\documentclass{aa}

\usepackage{natbib}
\bibpunct{(}{)}{;}{a}{}{,} % to follow the A&A style
\usepackage{epsfig}
\usepackage{graphicx}

\begin{document}

\title{The needle in the haystack - Where to look for more isolated cooling neutron stars}

%ICoNS in heavens

\titlerunning{ Where to look for new ICoNSs}

\author{B. Posselt
         \inst{1,2,3}
       \and
S.B. Popov
         \inst{4}
       \and
F. Haberl
         \inst{1}
       \and
J. Tr\"umper	 
         \inst{1}
       \and
R. Turolla
         \inst{5}
       \and
R. Neuh\"auser	 
         \inst{2}
       \and
P.A. Boldin	 
         \inst{6}
         }
\authorrunning{B. Posselt et al.}
   \offprints{B. Posselt}

\institute{Max-Planck-Institut
  f\"{u}r extraterrestrische Physik, Postfach 1312
85741 Garching, Germany 
         \and
Astrophysikalisches Institut und Universit\"{a}ts-Sternwarte, Schillerg\"{a}\ss chen 2-3,
07745 Jena, Germany
        \and
Observatoire Astronomique de Strasbourg, 11 rue de l' Universite, 67000 Strasbourg, France\\	
\email{bposselt@cfa.harvard.edu}
        \and
Sternberg Astronomical Institute,
Universitetski pr. 13, 119991 Moscow, Russia\\
\email{polar@sai.msu.ru}
        \and
University of Padua, Department of Physics, via Marzolo 8, 35131 Padova, Italy
        \and
Moscow Engineering Physics Institute (State University), Moscow, Russia
}
   \date{}

\abstract 
   {Isolated cooling neutron stars with thermal X-ray emission remain rarely detected objects despite many searches investigating the ROSAT data.}
   {We simulate the population of close-by young cooling neutron stars to explain the current observational results. Given the inhomogeneity of the neutron star distribution on the sky it is particularly interesting to identify promising sky regions with respect to on-going and future searches.}
   {Applying a population synthesis model the inhomogeneity of the progenitor distribution and the inhomogeneity of the X-ray absorbing interstellar medium are considered for the first time. The total number of observable neutron stars is derived with respect to ROSAT count rates. In addition, we present sky maps of neutron star locations and discuss age and distance distributions of the simulated neutron stars. Implications for future searches are discussed.}
   {With our advanced  model we can successfully explain the observed log~N~--~log~S distribution of close-by neutron stars. Cooling neutron stars will be most abundant in the directions of rich OB associations.
   New candidates are expected to be identified behind the Gould Belt, 
in particular in the  Cygnus-Cepheus region. 
They are expected to be on average  younger and then hotter than the known population of  isolated cooling neutron stars.
In addition, we propose to use data on runaway stars to search for more
radio-quiet cooling neutron stars.
}
   {}

\keywords{stars: evolution  --- stars: neutron --- X-rays: stars}

\maketitle
\section{Correction remark -- added March 2010}
Working on the new version of our population synthesis code,
we recently discovered a bug in the version used in Posselt et al. (2008).
This bug concerns the application of the improved ISM models to calculate the
absorbed X-ray flux. The code erroneously did not cover the whole
Galactic coordinate ($l$,$b$) range of the ISM-models to obtain the absorbing column density
$N(H)$. Instead, only a small region ( $\triangle l\approx 7\deg$, $\triangle b
\approx 7\deg$ ) around the Galactic Center was used, which led to a
significant overestimation of the absorption translating into 
an underestimation of the predicted neutron star number.

While all our main conclusions remain valid we add notes on
some details regarding the corrected results in the Appendix~\ref{Erratum}. Updated figures are shown in the Appendix~\ref{Erratum} as well. It is mentioned in boldface in the text when there is an updated result available.

We note that this is the only paper, where the presented results are influenced by the bug in our population synthesis code. Previous works do not use it, and subsequent works like, e.g., \citet{Popov2010} have already applied the corrected version of the population synthesis code.

\section{Introduction}
\label{intro}

More than 10 years after the discovery of its brightest member RX J1856-3754 (Walter et al.
1996), the small group of seven radio-quiet isolated neutron stars
(NSs) detected by the ROSAT satellite  have gained an important place in
the rich zoo of compact objects. Together with Geminga and several
close-by young radio pulsars, these objects form the local
population of cooling NSs. Studies of this group of sources
already provided a wealth of information on NSs physics (see e.g.
\citealt{Haberl2007,Page2007,Zane2007} for recent
reviews).

Since 2001 the number of known close-by radio-quiet NSs has not been growing
despite all attempts to identify (mainly in the ROSAT data) new candidates.
Partly this is due to the fact that all these searches are blind, i.e. not necessarily targeted to promising sky regions.
 To advance the identification of new near-by cooling NSs it is
necessary to perform a realistic modeling of this population.
The seven ROSAT radio-quiet NSs form the main part of the known local population of
cooling NSs.
For these seven sources we use the term {\it Magnificent Seven}.
There is still no general term to address the whole population to which
these sources belong (terms such as XDINS -- X-ray dim isolated NSs, DINS -- dim isolated NSs, RINS -- ROSAT (or radio quiet) isolated NSs,  are used sometimes by different authors). In this paper we use the term ICoNSs (Isolated Cooling NSs). In our opinion, this variant better reflects the nature of these sources than, for example, XDINS, as many of them are not dim in X-rays.
Finally, we call {\it coolers} all NSs, whose residual  thermal
emission can be detected.
Besides the radio-quiet NSs, such sources can be also known as normal radio pulsars, or they may demonstrate some kind of radio emission or even $\gamma$-ray emission distinct from classical pulsars.
In this study we do not care if an object shows some
activity in addition to thermal surface emission.
The only crucial point is the detection of X-ray emission related to
the cooling of initially hot compact objects.

The population synthesis (PS) approach is used here to investigate the population of close-by young cooling NSs.
There are two main versions of this method 
\citep{Fritze-v.Alvensleben00}. One, which sometimes is also called {\it
empirical population synthesis}, is a kind of top-down approach, when
objects, for which only integral characteristics are available, are studied.
Numerical models are applied to reproduce observed properties of such
sources by using different sets of subpopulations which constitute an
object class. For example, the stellar content of a galaxy can be derived by
modeling its integral spectra. 

Here, we focus on the second -- bottom-up --
approach, often called {\it evolutionary synthesis}. In this case the evolution
of a population of objects is modelled from their birth based on defined
initial distributions of parameters and some equations which describe
evolution of these distributions in time. The main reasons to use this
second technique
are the following. At first, if initial parameters or/and evolutionary laws for
some kind of sources are not well known, then this kind of PS  can be applied to test
hypotheses about these uncertain properties by comparison between modeled
and observed populations. Determination of the initial parameters of radio
pulsars can be a good example here \citep{faucher2006}. Next, even if the initial distributions and
their evolution are known, but at the moment just a small part of the
population -- a tip of an iceberg -- is observed, then PS calculations can
be used to predict properties of the unobserved faint part of the population, and
to plan a search strategy to identify new members. 
A more detailed review of the PS technique can be found, for
example, in
\citet{pp2005}.

In this paper an upgraded population synthesis model is discussed for the population
of close-by ($< 3$~kpc) isolated NSs which can be
observed via their thermal emission in soft X-rays.
Previously, our models were applied to confirm the link between the Magnificent Seven and the
Gould Belt \citep{p03}
%(Paper I)\nocite{p03},
and to test theories of thermal evolution of NSs \citep{pgtb04}.
%(Paper III) \nocite{pgtb04}.
The major interest of the present study is to understand how to find more objects of this type.
After  the description of the PS model ingredients in Sec.~\ref{model},
 we present our results -- $\log$~N~-~$\log$~S-curves, age and
distance diagrams as well as sky maps of expected NS locations -- in Sec.~\ref{results},
and discuss possible applications in Sec.~\ref{search}. An outlook concerning the search for ICoNSs is given in Sec.~\ref{outlook}
Finally, we give our conclusions in Sec.~\ref{concl}.

\section{The new population synthesis model}
\label{model}
The main physical ingredients which constitute our population synthesis model
are the following:

\begin{itemize}
\item[(A)] the initial NS spatial distribution and the NS birth rate;
\item[(B)] the kick velocity distribution of the NSs;
\item[(C)] the Galactic gravitational potential;
\item[(D)] the distribution of NS masses, in the following: NS mass spectrum;
\item[(E)] the NS cooling curve;
\item[(F)] the NS surface emission in X-rays;
\item[(G)] the interstellar absorption of X-rays;
\item[(H)] the properties of the X-ray detector.
\end{itemize}

At each time step in the simulation we consider at the same time eight different NS masses with corresponding cooling curves, following each NS on its trajectory through space and time until it is too faint to be observed.
The overall result of, e.g. $50000 \times 8$ simulated NS evolutionary
tracks from birth till the time when the temperatures falls below $10^5$~K, 
is normalised by the mass distribution as well as by birth rates (see below for both). 
Ingredients B, C, E, and F
are unchanged with respect to our previous studies (\citealt{p03, p04, pgtb04}, hereafter Paper I, II, III), so
we just briefly comment on them below.
The ingredients A, D, G, and H -- the initial spatial distribution,
the NS mass spectrum, the interstellar absorption,
and the X-ray detector properties -- are modified, and they
are described in subsections below in more details.

For the kick velocity distribution (B) of newborn neutron stars we use that proposed by \cite{acc2002}, as in all our previous studies.
This is a bimodal distribution, consisting of two Maxwellians with most
probable velocities
$V_{\mathrm {peak1}}\approx 127$~km~s$^{-1}$ and $V_{\mathrm {peak2}}\approx
707$~km~s$^{-1}$, the average velocity is $\approx 527 $~km~s$^{-1}$.
Recently, this assumption was criticized. A  single-component distribution with  $V_{\mathrm {peak}}\approx 400$~km~s$^{-1}$ was
advocated by \cite{h2005}, and by \cite{faucher2006} who accepted a non-Maxwellian velocity distribution.
However, since all objects under study are very young
and so still are relatively close to their birth place, the exact shape of the
velocity distribution does not have a strong influence on
our final results.

The calculation of NS trajectories
in the Galactic
gravitational potential (C) follows that presented in
Papers I and II \nocite{p03, p04}. As before we use the axisymmetric
Miyamoto-Nagai (1975) \nocite{mn1975} potential with
disc, buldge and halo contributions. This potential (with some
modifications) was actively used in NS calculations by different authors
already in the 1990s (see, for example, \cite{bm1993} and references therein).
Again, it is noted that since the NS are young the use of a very precise multicomponent potential is not
necessary.
Thus, we neglect also the Galactic bar which extends to a maximum Galactocentric
distance of 3.5~kpc \citep[e.g.][]{Pichardo2004}
because its influence is very small for the relatively nearby
objecs of interest.

The thermal evolution of NSs is one of the main ingredients of the
study. There are strong uncertainties in this topic. Sets of curves have
been calculated by  many research groups for different
models of NS interiors (see a reviews, e.g. by \citealt{page2006} or \citealt{Yakovlev2004}).
In the following we consider only hadron stars
and use cooling curves (E) corresponding to those marked as model~I in
Paper III \nocite{pgtb04}.
This cooling model includes the following important ingredients.
Superfluid gaps are taken according to \cite{tt2004}.
Medium modifications of the neutrino processes are taken into account.
Formation of the pion condensate is possible.
The relation between temperatures of the outermost core layer and the surface
is taken from \cite{bgv2004}. No additional heating is used.

In our calculations we do not account for atmospheric
reprocessing of the thermal radiation, and assume that the emitted spectrum  is a
pure blackbody (see a detailed description
of the modeling of the observed neutron star thermal emission in
\citealt{zavlin2007}).
Unfortunately, the distribution of the chemical compositions of atmospheres is not known
for the whole population of young NSs. 
If the atmospheric composition changes in time or not is also unclear.
As an additional parameter one needs further to consider
the (unknown) magnetic field strength and its effects on the atmosphere. 
So, the available data are not complete enough and statistics is too low to consider atmospheres in population synthesis studies.

However, isotropic blackbody emission is a simplification since there is
growing evidence that the thermal X-ray emission observed in case of the Magnificent Seven
actually comes from hot spots and not from the whole NS surface (e.g.
\citealt{Haberl2006,Schwope2005,truemper2004}). 
This is expected to influence not only the cooling times but also the detection probabilities.
But the overall quantitative effect is currently unclear as at first it is necessary to include non-uniform heat transport in the outer NS layers into cooling curves calculations self-consistently, for example, in the way it is discussed in \cite{apm2007}.
Strong toroidal magnetic fields produce hot spots and can keep them hot over a longer time than
the usual lifetime expected from the previous cooling models
\citep{Page2007}. 

Since the hot spot of a neutron star will
have a higher temperature than in the case of isotropic cooling, its
detection would be less affected by interstellar absorption. The fact that
these effects are not accounted for by our applied cooling curves leads to 
underestimating the number of observable neutron stars. On the other hand,
\citet{Yakovlev2004} have pointed out that at high
magnetic field strength the
break of the cooling curves (marking the transition from neutrino to photon
cooling) shifts towards lower ages, viz. from $\sim 10^6$ years at B$\sim 0$\,G to a few
times $10^5$ years for magnetic fields typical for the Magnificent Seven,
B$\sim 10^{13}$ - $10^{14}$\,G.  
Neglecting this effect will lead to an overestimate of the
number of observable neutron stars with ages of (0.5-1)$\times 10^6$ years where some
of the Magnificent Seven are found. 
Magnetic field decay and the additional heating due to it \citep{Aguilera2007astroph} can
further increase the number of observable NSs.
We cannot make a firm prediction about the net
change to the $\log{N}$ - $\log{S}$ which is caused by these effects which are partly
counterbalancing. However, we believe that the blackbody estimate of the
count rate $S$ is reliable within a factor of $\sim 2$.

The soft X-ray spectra of objects under consideration are well fitted by the blackbody shape, and, more important, the
cooling curves used here were calculated under the assumption of blackbody emission 
of the whole NS surface, and cooling curves for non-uniform
emission are currently not publicly available.
We therefore stick to the cooling curves of paper III and the assumption of spherical symmetric blackbody emission, and apply them as a reference model to study the main aspect of this paper -
the influence of the inhomogeneous distributions of neutron star birth places
and interstellar absorptions in the Galaxy.

\subsection{Initial spatial distribution of NSs}
\label{ini}

Following the results of
previous investigations (paper~I\nocite{p03}), we take as
established that the population of nearby NSs is genetically
related to the Gould Belt. 
The
contribution of the Gould Belt dominated the production of compact
remnants in the solar proximity over the past $\sim 30$~Myrs (see
\citealt{p97} for a detailed description of the Gould Belt structure).
About two thirds of massive stars in the $\sim 600$~pc around the
Sun belong to the Gould Belt (\citealt{torra}). 

In  paper I and
II\nocite{p03, p04} we considered that NSs are born either in the
Belt or in the Galactic disc, both treated as infinitesimally thin,
disc-like structures, with a uniform spatial distribution. The
total NS birth rate in the Galactic disc (the whole area inside 3 kpc) and that
in the Gould Belt alone were taken to be 270~Myr$^{-1}$ and 20~Myr$^{-1}$, 
respectively. 
The first value corresponds to the value inside 1 kpc by \cite{tammann}. 
The
latter value is from \cite{g2000}, and appears in reasonable
agreement with the historical supernova rate inside 1 kpc, as
estimated by \cite{tammann}.

Here, the previously simplified progenitor
distribution is significantly upgraded by adopting a more realistic description of the massive star
geography in the solar neighborhood. In particular, in the innermost 500 pc
from the Sun the initial spatial distribution of NSs is assumed to follow the
presently observed distribution of massive stars. Outside the well-known 500 pc and up to 3
kpc, NSs are born either 
in the Galactic disc, assumed to be exponential both in $z$ and $R$
directions, 
or originate from known, rich OB associations.

\subsubsection{Inside 500 pc}

For the stars inside 500 pc we use the {\it Hipparcos} data on massive
stars \citep{esa97}. Our assumption is that the present day distribution of massive stars represents their distribution in the last 1-2 million years well.

All the 570 classified B2-O8 stars with
known parallaxes $>0\arcsec.002$ are considered.
According to the birthrates above in our model 27 (out of 270) NS producing 
supernova events per Myr are assumed
to be related to this population. This number 
should be compared
with 20 NSs assumed to be born in the entire Gould Belt in 1 Myr in our
old model. The parameters of the new model roughly lie in between
the old one for $R_{\mathrm{Belt}}=$~300 and 500 pc.

\begin{table}[b]
\begin{tabular}{|l|c|}
\hline
  OB association & Age [Myrs]\\
\hline
 Upper Scorpius & 4-6\\
 Upper Centaurus Lupus & 12-14\\
 Lower Centaurus Crux & 9-11\\
 Vel OB2 & 6-10\\
 Trumpler 10 & 15-30\\
 Collinder 121 & 4-6\\
 Ori OB1  & 9.5-13.3\\
Per OB2 & 3-7\\
 $\alpha$ Persei (Per OB3) & 25-50\\
 Cas-Tau  & 25-50\\
 Lac OB1 & 2-25\\
 Cep OB2      & 5-10\\
 Cep OB6  & 25-50\\
 \hline
\end{tabular}
\caption[]{OB associations inside 500 pc and their ages (from de Zeeuw et
al. 1999).
For Ori OB1 the age proposed for its older part Ori OB1A is taken.
}
\label{tab:OBage}
\end{table}

In the code the initial position of each newborn NS coming out of
this population coincides with the position of one of these 570
{\it Hipparcos} stars. This assumes that the spatial distribution
of the massive main sequence stars does not change significantly
over the NS evolutionary timescale ($\approx 1$ Myr). 
A progenitor star is chosen randomly, but we tried to account for
the probability for each particular star to explode. Since for
each star the spectral type is known, one can roughly estimate its
mass. The mass interval corresponding to each spectral type
 gives us an estimated range for its total lifetime. If one has no
information about the age of the star then the probability is assumed to depend just
on the star's full lifetime: the probability is higher for more
massive stars as they are shorter lived (actually, the probability to explode is
inversely proportional to the lifetime of the star).

\begin{figure*}[t]
\vspace{-0.2cm}
\vbox{\psfig{figure=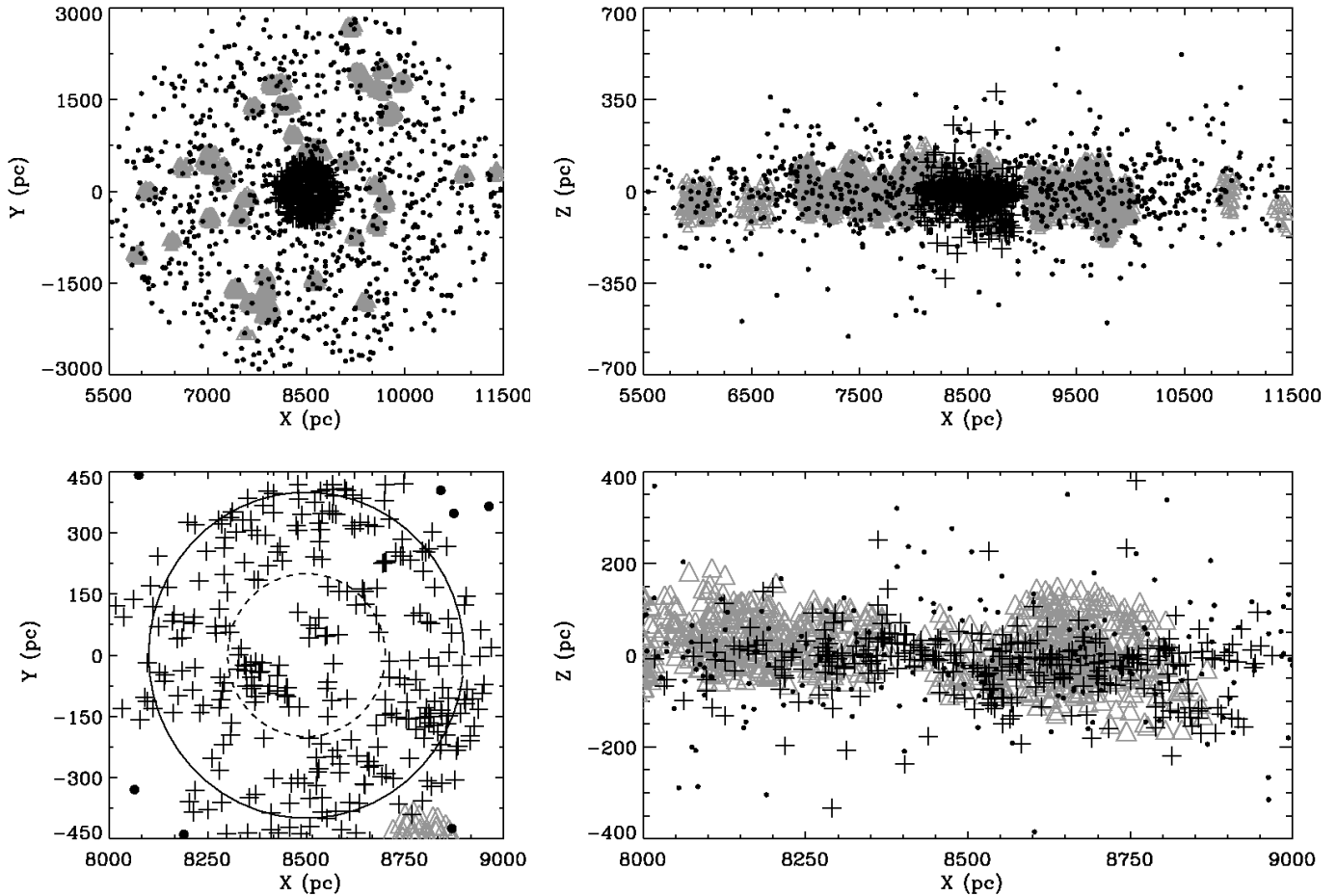,width=\hsize}}
\vspace{0.19cm}
\caption[]{The new NS progenitor distribution.
The grey triangles represent the stars in the OB associations. Crosses stand for the considered
Hipparcos stars inside 500
pc. Small circles mark random stars in the Galactic disc ($500<d<3000$~pc).
Outside 500 pc from the Sun two populations are visible: NSs formed in OB associations, and NSs formed in the field.
Bottom panels show the projected distribution X-zoomed to 500 pc around the Sun.
Two circles on the bottom left panel
are plotted with radii of 200 and 400 pc from the Sun.
}
\label{fig:xyz_ini}
\end{figure*}

As noted above, the mass interval corresponding to each spectral class is
converted into an approximate limiting lifetime:  T$_{LM}$ is the
longest lifetime for the lowest mass, and T$_{HM}$ is the shortest
lifetime for the high mass end of the particular mass interval. If
the star is known to belong to an OB association, we also consider
the age of the OB association. The ages of the OB associations are shown in
Table~\ref{tab:OBage}; they are taken from \citet{Zeeuw1999}. In several cases \citet{Zeeuw1999} give just one number (not a range) as an estimate of an age. Of course, these numbers have some uncertainty. To account for this we added an uncertainty
($\pm 1$ Myr) 
in all those cases.
By this we have for all OB associations in the Table TL$_{OB}$ and TU$_{OB}$, a lower and upper age
limit for the OB association, respectively. 

The probability for a
progenitor star to explode (and so to contribute to NS production in our model) is estimated as follows.
\begin{itemize}
\item[(i)] If $T_{HM}>TU_{OB}$, then the probability of a star
exploding is zero, as the shortest estimated lifetime is longer
than the age of the OB association. Such stars do not contribute to NS production in our model, i.e. no compact objects could have been born at their locations because the OB association is too young.

\item[(ii)] Conversely, if $T_{LM}<TL_{OB}$ then the probability
is inversely proportional to the longest lifetime, T$_{LM}$, of
the star. This probability is also used for stars not belonging to
OB associations within a 500~pc distance.

\item[(iii)] If $T_{HM}<TL_{OB}$ but $T_{LM}>TU_{OB}$ then the
probability is inversely proportional to $(TU_{OB}-TL_{OB})$, as
one has to consider overlapping time scales.

\item[(iv)] If $T_{HM}<TL_{OB}$ and $TL_{OB}<T_{LM}<TU_{OB}$  then
the probability of exploding is inversely proportional to
$(T_{LM}-TL_{OB})$.

\item[(v)] If $TL_{OB}<T_{HM}<TU_{OB}$ and $T_{LM}>TU_{OB}$  then
the probability is inversely proportional to $(TU_{OB}-T_{HM})$.

\item[(vi)] If $TL_{OB}<T_{HM}$, but $T_{LM}<TU_{OB}$ then the
probability of exploding is inversely proportional to the minimum
time value $(T_{LM}-T_{HM})_{MIN}=1.44$~Myr, which was obtained for 
the spectral type B0 corresponding to the mass
interval  $[ 15$~M$_\odot , 18$~M$_\odot ]$.
\end{itemize}

To summarize, the initial spatial distribution of NSs inside 500 pc in our model follows the distribution of {\it Hipparcos} massive stars weighted by their explosion probabilities as estimated above. Outside 500 pc this weighting is not applied.

\subsubsection{Outside 500 pc, up to 3 kpc}

Most of the NSs in our model, 200 out of 270 per Myr, 
are born in one of 49 OB associations
outside 500~pc and up to 3~kpc.
The radii of the association extents are taken to be equal to 100
pc. This is a safe upper limit on their sizes, as usually
diameters are about 80 pc, and up to 200 pc \citep{me1995}.
The probability to be born inside this radius is uniform. 
We assume that the number of stars with known photometry roughly
reflects the total number of stars in the cluster. The probability
to be born in each cluster is proportional to this quantity. The
data on the 49 OB associations are taken from the catalogue by
\cite{bh1989} (see also \citealt{me1995}). Note, that distances to
the associations are reduced by 20\% according to conclusions by
\cite{d2001}.

Finally, out of the total 270 NSs, 43 are supposed to be born
during 1 Myr in the Galactic disc outside 500 pc from the Sun, but
inside 3 kpc. Their positions in the disc are calculated randomly.
The disc is exponential both in $R$ and $Z$. For the disc Z-scale
we take the value 100 pc. For the radial scale we take the value 3
kpc (\citealt{kent1991,fux1994}).

In our model the NS formation rate outside 500 pc around the Sun
corresponds to $\sim 0.884 \, 10^{-11}\, {\rm pc}^{-2}\, {\rm yr}^{-1}$.
 This rate is lower than the one stated by \citet{tammann}. 
They, adopting that the Milky Way is type Sbc, and
combining the corresponding value with the historical rate in the Galaxy, reported a local SN rate of  
$2 \, 10^{-11}\, {\rm pc}^{-2}\, {\rm yr}^{-1}$.
\citet{tammann} gave also a rate of $2.9 \, 10^{-11}\, {\rm pc}^{-2}\, {\rm yr}^{-1}$ which is based on the census of O-B2 stars in 1 kpc around the Sun. Clearly, both values should be dominated by the productive Gould Belt. Outside the Belt the rate should be lower, however, it should be anisotropic and non-uniform on the scale of 3 kpc.
Our adopted values, $\sim 3.438 \, 10^{-11}\, {\rm pc}^{-2}\, {\rm yr}^{-1}$ inside 500~pc and $\sim 0.884 \, 10^{-11}\, {\rm pc}^{-2}\, {\rm yr}^{-1}$ outside 500~pc are chosen to reflect this fact.
At the 500~pc transition our model gives comparable values due to the scaling through OB associations. We checked that at this transition our model distance distribution of {\it unabsorbed} NS show consistently increasing numbers.
Probably, we slightly underestimate  the NS formation rate at larger
distances in the directions of the Galactic center and spiral
arms.
However, for our calculations closer regions are more important (see
distance distributions of ICoNSs below).
Anyway, we warn the reader that some uncertainty in the birth rate
normalization is hidden inside our model.
Comparison of the Log N -- Log S distribution for old and new
models of the initial spatial distribution of NSs is made in
Fig.~\ref{fig:ini_xyz} and discussed in Sec.~\ref{results}.

\subsection{The interstellar absorption of X-rays}
\label{ism} X-rays are absorbed by the heavy
elements of the interstellar medium. This is especially true for
the soft X-rays below 1~keV. Their extinction has to be accounted
for if one aims at realistic estimates of the sources
observability. The amount of material between the object and the
observer is measured as hydrogen column density $N(H)$. Then,
assuming the same abundances towards all directions and accounting
for photoelectric absorption processes by atoms, ions and dust
particles one can determine the X-ray extinction. 
The treatment of X-ray absorption is significantly upgraded in our new population synthesis with respect to
previous studies by considering all these aspects , i.e. a more
realistic model for $N(H)$, up-to-date abundance tables and cross
sections (see \citet{Posselt2006} for a more thorough discussion
and paper III for details about previous treatment of the ISM
absorption\footnote[1]{We also corrected in the old ISM model the use of full widths of half maximum instead of the required variances in the exponential density factors by \citet{Zane95,Dickey90}. However, the effect of smaller ISM density above the Galactic plane on our model results was quite low. }
). 

In the following we use the abundance tables and photoelectric
cross sections by \cite{Wilms2000} as combined in the
\texttt{tbabs} routine which is implemented e.g. in
XSPEC\footnote[2]{http://heasarc.gsfc.nasa.gov/docs/xanadu/xspec/}.
We also introduce two different, new $N(H)$ models in three
dimensions to account for the highly inhomogeneous distribution of
the interstellar medium (ISM). Both models take into account the
most updated results in the local neighborhood (up to 230~pc) by
\citet{Lall03}. Their work well describes the local bubble which
is  surrounded by clumps of denser
ISM and connected by chimneys to other bubbles. 
For distances larger than 230~pc we
use the analytical ISM model described in \citet{p00b} in one of our new $N(H)$ models, in the following it is named ``analytical
ISM model''. For the second $N(H)$ model we use the extinction
study by \cite{Hakkila97}. In the following we refer to this model
as the ``Hakkila ISM model''. 
 
Concerning the transition region at 230~pc we want to make the following short comments.
The column density can only increase with larger distance, thus the new data from the analytical formulae / the Hakkila study are only applied if the resulting column densities are larger than those from \citet{Lall03} at 230~pc.
At distances larger than 230~pc the models become naturally much coarser due to fewer measurements. 
For the analytical model the transition is smooth since the analytical formulae at, e.g., 240 pc result in values similar to the ones at 230~pc by \citet{Lall03}.
In case of the Hakkila model the transition is  more abrupt due to the patchiness of the available measurement data applied for the extinction study by \cite{Hakkila97}.
On $average$ the hydrogen colum density at around 230~pc is around $10^{20}$~cm$^{-2}$. This $average$ value 
increases by around $10^{20}$~cm$^{-2}$  at the transition for the Hakkila model, while in case of the analytical model the $average$ increase is only around $10^{19}$~cm$^{-2}$. 

Both models were checked with open
cluster extinction data which were presented by
\citet{Khar2005b,Khar2005a} and \citet{Piskunov2006} and are complete up to a
distance of 850~pc. The data of the open clusters were then
included in both models (see \citet{Posselt2006} for details on the models).
The
column densities derived from both ISM models at large distances
were also compared  with the results by \citet{Schlegel1998} from
infrared dust emission measurements. In both cases, the analytical
ISM model and the Hakkila ISM model, data cubes in Galactic
coordinates (steps of one degree) and distance (steps of
10~pc) are used for our population synthesis. The sampling of the
data cube does not reflect the actual accuracy, which is around
25~pc within 230~pc and much coarser at larger distances.

It is noted that the both models are large-scale models and even the
Hakkila ISM model does not account for small ISM clumps,
especially at large distances, above 1~kpc. 

\subsection{Mass spectrum}
\label{mass}

NS cooling curves strongly depend on masses. 
Once the cooling scenario is chosen, it is
therefore necessary to specify the NS's mass spectrum. 
As in our previous studies, we use the mass spectrum of NS which is derived
by the joint use of {\it Hipparcos} data on spectral classes of
close-by massive stars and calculations by \cite{whw02}. In addition, in this paper a modified mass spectrum is studied obtained by using
the  calculations from \cite{hws2005}.

Both mass spectra are obtained in a similar way, and {\it a priori} it is impossible to say that the ``new'' one is preferable with respect to the ``old'' one. We take spectral
types of massive stars around the Sun as given in the
{\it Hipparcos} catalogue \citep{esa97}. To each spectral type we
associate a range of masses. Applying now the calculations by
\cite{whw02} or, instead, by \cite{hws2005}
we obtain the baryonic masses of compact
objects. Then the baryonic masses are translated into
gravitational masses by using the simple equation, valid for
hadronic NSs, by \citep{tww1996}

$$
M_{\mathrm{bar}}-M_{\mathrm{grav}}=0.075 \, M_{\mathrm{grav}}^2.
$$
The accuracy of this relation is sufficient for our purposes, as
one needs to know the mass within a few percent (less than
half the size of a mass bin). Finally, we have a mass spectrum
with eight mass bins which correspond to the eight cooling curves
we use in our calculations.

\begin{figure}[t]
\vbox{\psfig{figure=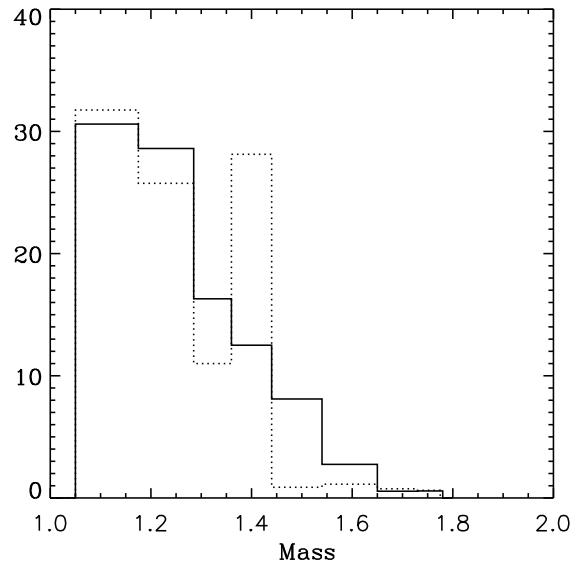,width=\hsize}}
\caption[]{The old (as used in in previous studies, dotted line) and the new (as  introduced in this work) mass spectra, 
binned over eight intervals of
different widths (see the text).}
\label{fig:mass2}
\end{figure}

In Fig.~\ref{fig:mass2}  
the used two mass spectra are shown --
the old (dotted line in the graph) mass spectrum which we
also used in previous studies, and the new one. 
Note the non-equal widths of the
mass bins. Eight cooling curves for gravitational masses 1.1,
1.25, 1.32, 1.4, 1.48, 1.6, 1.7, and 1.76~$M_\odot$ are taken, both in the new and old mass spectrum.
Each bin
boundary corresponds to the mean between two values of mass used
in calculations.

For the old 
spectrum in Fig.~\ref{fig:mass2} we use the relation between
a progenitor mass and NS baryonic mass derived from Fig.~17
by~\citet{whw02}. We approximate the data in 
this figure with two linear relations (for progenitor masses $<15
\, M_\odot$ and $>25\, M_\odot$, respectively), and constant
between 15 and 25 solar masses (see Paper II). A peak 
at 1.4~$M_\odot$ appears due to the fact that in
this region progenitors of different mass produce NSs of similar
masses. The new mass spectrum, shown with a solid line,
is derived much in the same way, but using
data from \cite{hws2005} for stars with masses $>12\, M_\odot$. As
it is visible in the figure, the peak at $1.4\, M_\odot$
disappears. This is because the dependence of the compact star
mass on the progenitor mass contains no flat regions.
The most important feature of both mass spectra is a very small
number of NSs with masses above $\sim 1.5\, M_\odot$. As the
cooling of a NS is faster for massive stars, a small number of
massive objects implies a small number of cold stars in the same
age group.

\begin{figure}[t]
\vbox{\psfig{figure=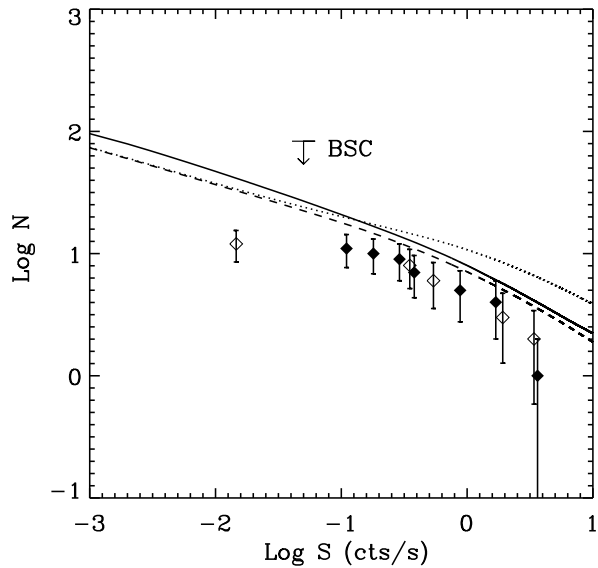,width=\hsize}}
\caption[]{Log N~--~Log S for the new initial spatial distribution (solid line) and two
variants of the old spatial distribution with $R_{\rm{belt}}=300$~pc (dotted) and 500~pc
(dashed).  The curve shown as a solid line is used as reference on
the following plots, too. This curve is calculated for the old mass
spectrum, old ISM distribution, and new ISM element abundances. 
The data points represent the Log N - Log S distribution for 
the known young close-by 
isolated neutron stars with thermal X-ray emission (the
Magnificent Seven, Geminga, the ``second Geminga'' -- 3EG J1835+5918, and
three radio pulsars: B0833-45, B0656+14, B1055-52).  
Filled symbols correspond to the Magnificent Seven, open symbols to the other NSs.
Error bars are plotted for the poissonian errors. 
From the list in
\citet{p03} we removed PSR1929+10 since results of the recent XMM-$Newton$
observations by \citet{Becker2006} demonstrate that the spectrum is
non-thermal.
}
\label{fig:ini_xyz}
\end{figure}

\begin{figure}[t]
\vbox{\psfig{figure=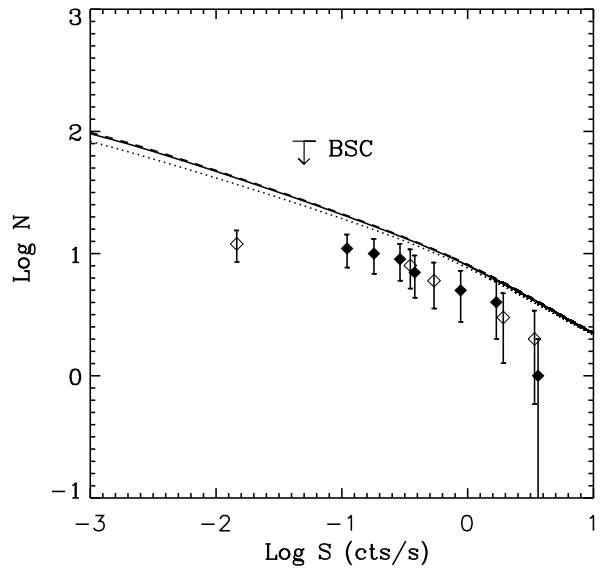,width=\hsize}}
\caption[]{In this figure we present the effect of the new mass spectrum
and new ISM element abundances. All curves are plotted for the new initial spatial
distribution of NSs.
Solid curve: old mass spectrum and new abundances. Dotted line: old mass
spectrum and old abundances. Dashed curve: new mass spectrum and new
abundances.   All curves are calculated for the old ISM distribution model.
}
\label{fig:mass_abund}
\end{figure}

\subsection{Detector properties}
\label{detec}

As before (e.g. 
in Paper I) we use the instrumental response of ROSAT to
draw conclusions with respect to the ROSAT All Sky Survey (RASS). However,
the energy range considered, e.g. in Paper I
was 0.08~keV to 3~keV while
the RASS data analysis was done for a slightly smaller energy range. To be
in full agreement we changed the energy range now to 0.1~keV to 2~keV. The
effective area of the instrument is discontinuous over energy due to 
 sharp instrumental absorption edges in the energy band of interest. With respect to previous studies
 our sampling of the effective area measurements was improved to  better account
for such absorption edges. In few cases we found differences of up to $40\%$
in the resulting absorbed fluxes for an object. However, this effect is pronounced only  for very soft sources at very small distances. Thus the influence
on the overall results of the population synthesis are small. We note, that
of course one has to consider vignetting of the X-rays as not all RASS sources are observed on-axis. This
is done by an effective area measurement with averaged vignetting applicable
for the RASS.

\section{Results of calculations and discussion
on the influence of model modifications}
\label{results}

In this section we compare results of different modifications of
the population synthesis scenario studied in this paper.
We start by comparing the  $\log$~N~--~$\log$~S distributions calculated with
different assumptions, proceed to the resulting map of the cooling
neutron star distribution on the sky,
followed by age and distance distribution, and
finally discuss the implications on searches for new ICoNS candidates.

\subsection{Comparison of $\log$~N~--~$\log$~S distributions for different model
modifications}
\label{lognlogs}

\begin{figure}[t]
\vbox{\psfig{figure=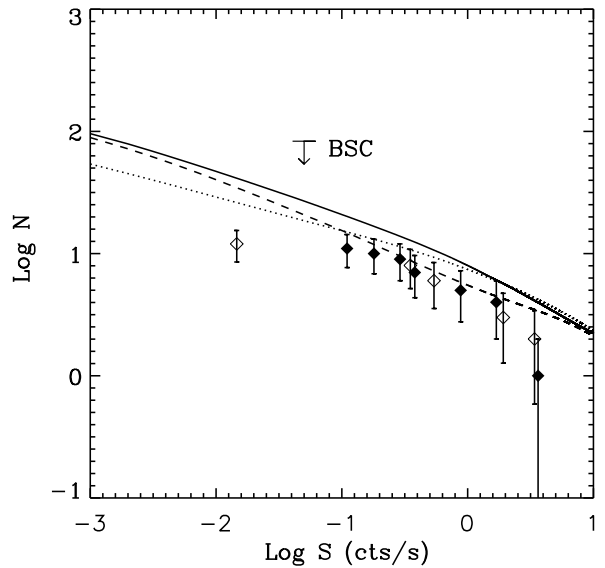,width=\hsize}}
\caption[]{Log N~--~Log S in case of different X-ray absorbing ISM models.
All curves are plotted for the new initial spatial distribution, the old mass spectrum and new ISM  element abundances.
Solid curve: old, simple analytical ISM model as e.g. in
Paper III \nocite{pgtb04}.
Dotted curve: new improved analytical ISM model;
dashed curve: Hakkila ISM model, see  Sec.~\ref{ism}
{\bf For an updated version of this figure see Fig.~\ref{fig:massISMupd} in the Appendix~\ref{Erratum}.}
}
\label{fig:massISM}
\end{figure}

The effects of the successive modifications in our PS-model on the $\log$~N~--~$\log$~S distribution are shown in Figs.\ref{fig:ini_xyz}-\ref{fig:massISM}. The solid curve in all the figures is the same. It corresponds to the new initial spatial distribution of the NSs, new abundances and cross-sections, the old mass spectrum, and the old ISM model.

At first we discuss the influence of the new initial spatial
distribution of NSs (i.e. distribution of progenitors).  We
compare three models: two variants of the old model with
$R_\mathrm{Belt}=300$~pc and 500~pc and the model with the new
initial spatial distribution described in Sec.~\ref{ini}. Log N --
Log S curves are shown in Fig.~\ref{fig:ini_xyz}. Except for the
initial spatial distribution all other parameters are the same: old
mass spectrum (Fig.~\ref{fig:mass2}, dotted curve), 
new abundances, old ISM
distribution (calculations with the new ISM distribution take
long times, so for comparison of different modifications we use the
old ISM distribution model when possible).

\begin{figure*}[th] 
 \centering 
 \includegraphics[width=0.9\hsize]{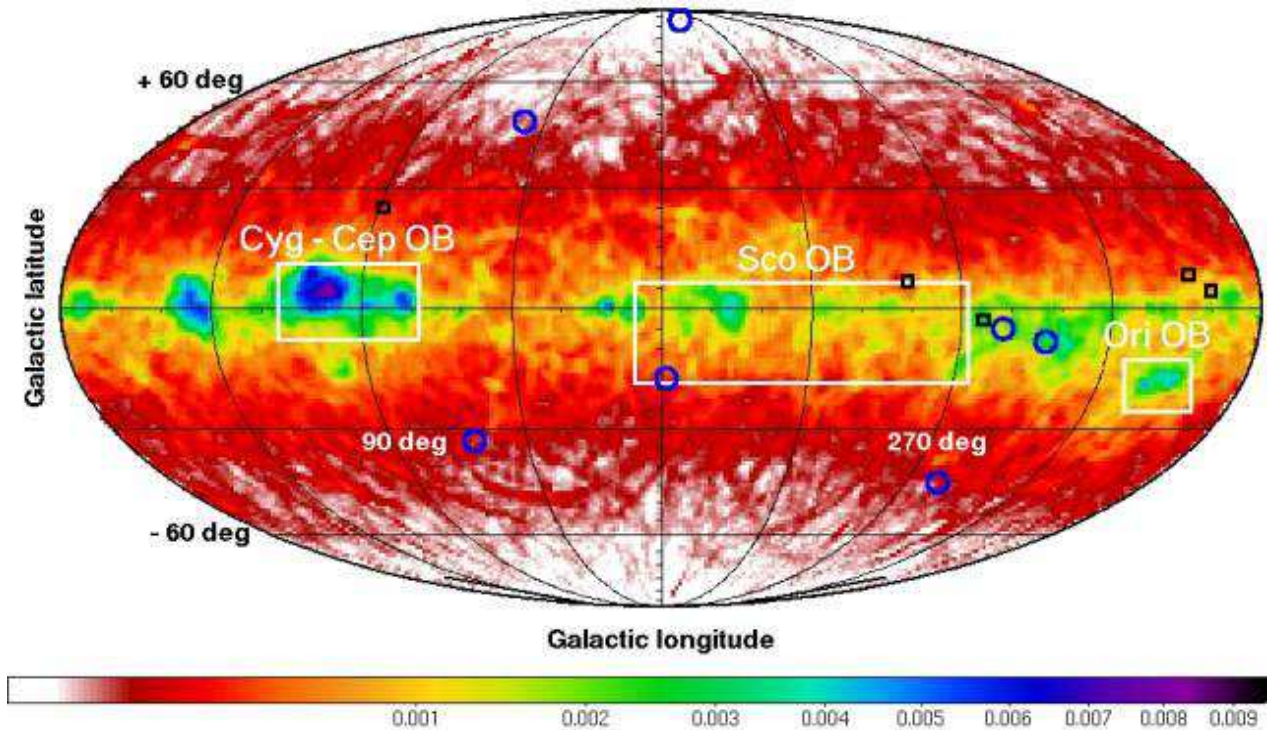}
\caption{The expected number density of isolated neutron stars
with thermal X-ray emission in units of numbers per square degree.
The Galactic map is in Mollweide projection.
Only sources with ROSAT PSPC count rates larger than 0.05~cts s$^{-1}$ are
considered, the same value as used in Paper II in their Fig.~6.
The simulation was done for new initial progenitor distribution,
new abundances, old mass spectrum and the new analytical ISM model,
thus corresponds
to the dotted $\log$~N~--~$\log$~S curve in Fig.~\ref{fig:massISM}.
Marked in blue are the positions of the Magnificent Seven and
in black the positions of close young radio pulsars with detected thermal X-ray emission.
{\bf For an updated version of this figure see Fig.~\ref{fig:mapErr} in the Appendix~\ref{Erratum}.}
}
\label{fig:map}
\end{figure*}

The solid curve corresponds to the new initial spatial distribution.
The result with the new progenitor distribution lies in between
those of the two old ones for large and moderate fluxes. In fact,
our new calculation is only slightly above the simpler model with
a Gould Belt radius of 500~pc. For count rates $\log S < -1$ the
new initial spatial distribution results in roughly $0.1$~dex more
observable sources. 
The new model of the spatial distribution and the old for $R_\mathrm{Belt}=500$~pc  are in reasonable correspondence
with observations at the bright end, where it is supposed that our
knowledge about observable objects is mostly complete (as we see below new models of the ISM distribution shift the curve down improving the correspondence). However,
the locations
of observable sources on the sky for the old and new initial spatial
distributions
are significantly different
(compare Fig.~\ref{fig:map} below and Fig.~6 in Paper II \nocite{p04}).

Next, one has to compare the Log N -- Log S distributions calculated for
different mass spectra (see Sec.~\ref{mass}).  Computations were made for
the new initial spatial NS distribution, new ISM element abundances and old ISM distribution.
Results are shown in Fig.~\ref{fig:mass_abund}.
To see the effect of the new mass spectrum one has to compare the solid and the dashed curves.
The solid curve is the same as in Fig.~\ref{fig:ini_xyz}. It corresponds to the
old mass spectrum (Fig.~\ref{fig:mass2}, dotted line). 
The dashed curve, which is nearly
indistinguishable from the solid one, is calculated for the new mass
spectrum (Fig.~\ref{fig:mass2}, solid line).  
Surprisingly, the two different mass spectra
provide nearly the same result for the Log N -- Log S. 
This is very positive for our approach since the mass spectrum is a poorly determined component of the model.
As noted above, none of the mass spectra applied here is currently more preferable than the other. In the following we continue to use the ``old'' mass spectrum.

In  Fig.~\ref{fig:mass_abund} the effect of the new ISM element 
abundances is  also shown. The dotted curve should be compared with the solid
one. As it can be  seen, the effect is not very important.
Naturally, it more strongly influences faint sources, which are
situated at larger distances (on average). However, even for
far away sources  the effect is not significant.

 Finally, we compare the $\log$~N~--~$\log$~S curves obtained for three different
models of the ISM distribution. In Fig.~\ref{fig:massISM} we show
results for the old simple model (reference curve), the new
analytical ISM model, and the Hakkila ISM model. The curve of the
new analytical ISM model is very close to the one of the old ISM
model for bright sources. As sources become fainter the number of
observable objects becomes smaller than before due to the appearance
of directions with high absorption. Using the Hakkila ISM model we
obtain a $\log$~N~--~$\log$~S curve which fits the known observational constraints even better. For small fluxes this
curve approaches the reference one.
In comparison to the analytical ISM cube there are large
``windows'' of low extinction in the Hakkila ISM model resulting
in more sources at low fluxes.  The analytical model on the other
hand seems to overpredict the absorption when going to larger
distances \citep{Posselt2006}. Thus, at fainter fluxes one may
regard the results of the analytical and Hakkila ISM model as
lower and upper limits on the number of observable sources.

We conclude, that in terms of the $\log$~N~--~$\log$~S distribution our new, more
realistic, models are in good agreement with the older ones. The
interstellar absorption by the Hakkila ISM model results in a $\log$~N~--~$\log$~S
curve situated slightly closer to the known observations of INSs.

\begin{figure*}[th] 
 \centering 
 \includegraphics[width=0.9\hsize]{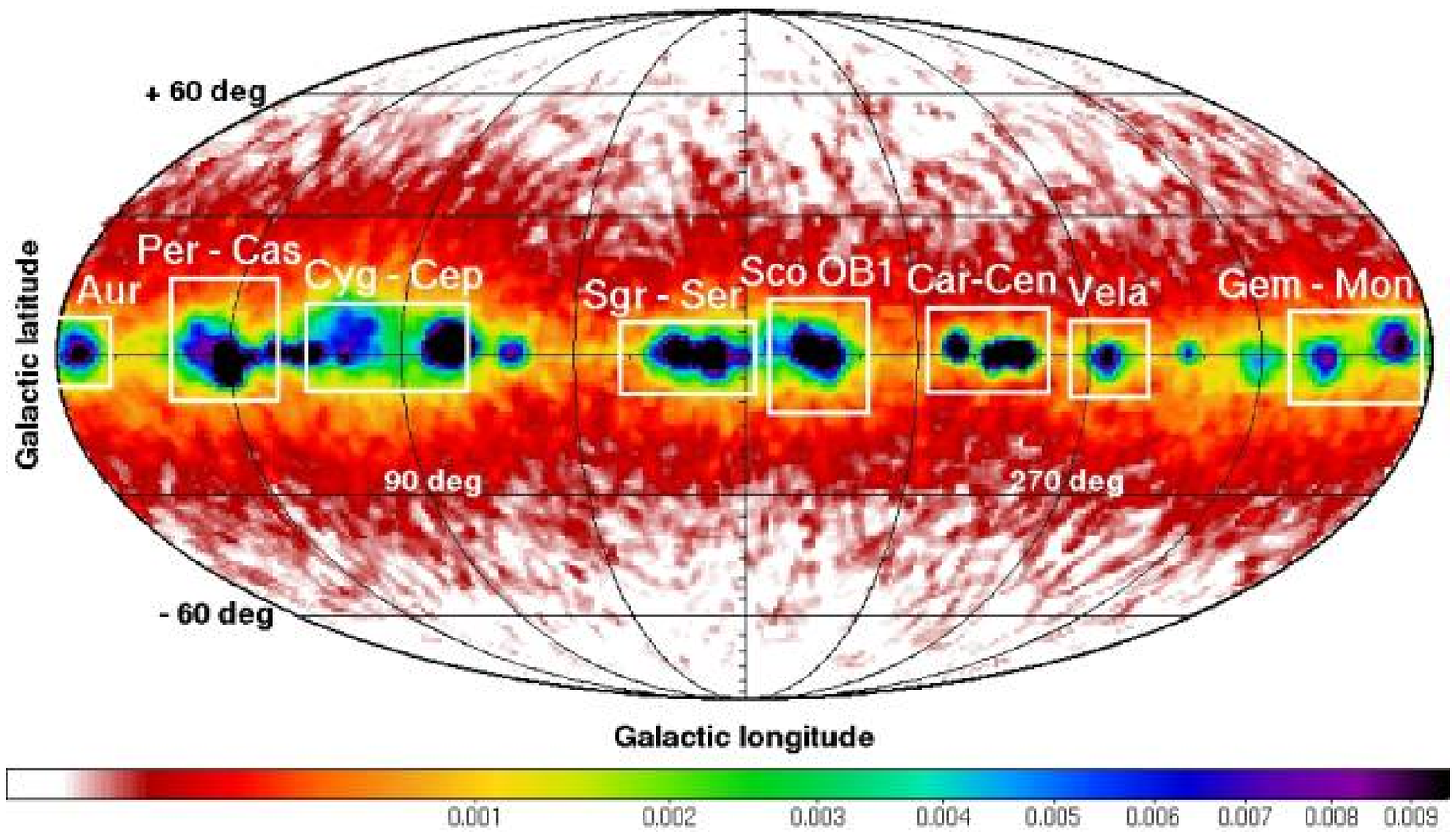}
\caption{Same as Fig.~\ref{fig:map}, but here only faint sources with ROSAT PSPC count rates between 0.001~cts s$^{-1}$ and  0.01~cts s$^{-1}$ are
considered.
Marked are regions of the corresponding OB associations giving birth to the neutron stars.
These OB associations have distances between 1~kpc and 2~kpc.
{\bf For an updated version of this figure see Fig.~\ref{fig:map2Err} in the Appendix~\ref{Erratum}.}
}
\label{fig:map2}
\end{figure*}

By using a different set of cooling curves one may improve the
agreement between computed and observed $\log$~N~--~$\log$~S distributions.
A detailed analysis is, however, outside the scope of the present
investigation.
Therefore, in the following only few comments are made
regarding the set of cooling curves applied in our previous Paper III.
As apparent from Fig.~\ref{fig:ini_xyz} and Fig.~\ref{fig:massISM} our final
$\log$~N~--~$\log$~S curves which include the new ISM distribution models
are not significantly different from the curve using
$R_\mathrm{belt}=500$~pc in our old calculations (Paper III).
So, the main conclusions about different sets of cooling curves presented in
Paper III should not change.
While the curve calculated in the case of the Hakkila ISM model -- dashed in
Fig.~\ref{fig:massISM} --
lies at low count rates slightly below the old curve with
$R_\mathrm{belt}=500$, this deviation is not strong enough to validate the
cooling models which have been rejected in Paper III.
Moreover, for this choice of the ISM distribution the cooling models which
have been considered as acceptable in Paper III fit even better the
observations.

\subsection{Sky maps}
\label{maps}

In earlier papers on population
synthesis calculations we did not attempt to produce realistic sky
maps of the ICoNS distribution, as some of our
assumptions were quite crude. Now, with detailed distributions of
ISM and birth places of NSs, one can try to compute such maps. 
In Fig.~\ref{fig:map} we present a map in Galactic coordinates for
the new initial spatial distribution of NSs, old mass spectrum and the new
analytical ISM model. The plot shows the expected number density
for sources with count rates $>$0.05~cts~s$^{-1}$. Obviously,
sources appear to be restricted towards the Galactic plane and
the plane of the Gould Belt. Few objects are expected to be found
at latitudes higher than 30$^\circ$ (see, however,
Sec.~\ref{outlook}). Inside $\pm$~30 degrees from the Galactic
plane the distribution of sources is dominated by NSs from
relatively close, rich OB associations. We mark by boxes the most
important of them: Sco OB2, Cyg OB7, Cep OB3, and Ori OB1.
Interplay between source distribution and 3D ISM structure allows
us to make predictions on which directions are most promising for
looking for new ICoNSs.

The computed distribution of sources in Fig.~\ref{fig:map} is dominated by sources with count rates in the
range $[0.05, 0.1]$~cts~s$^{-1}$.
This corresponds to the dimmest sources among
identified radio and $\gamma$-ray silent ICoNS, or even to count rates smaller than $<0.1$~cts~s$^{-1}$ where there aren't currently any
ICoNS known. In the distribution of known
ICoNSs \citep{Motch2007}
there are no sources with Galactic longitude in the range
$\sim50^\circ$ to $\sim 200^\circ$. At the first glance, this is
in contradiction with the our present map. However, it is
necessary to note, that for brighter sources ($>0.5$~cts~s$^{-1}$)
the Cygnus-Cepheus region does not give a strong contribution
, and 1-2 objects could  escape
identification in this overcrowded area close to the Galactic
plane (see  Sec.~\ref{search} for a more detailed discussion).

In Fig.~\ref{fig:map2} we present a similar map for the same parameters concentrated on the faint sources
having nominal ROSAT count rates between $0.001$~cts~s$^{-1}$ and $0.01$~cts~s$^{-1}$.
We note that the RASS is not so deep. Fig.~\ref{fig:map2} demonstrates the effect of more distant ($>1$~kpc) OB associations, the most important ones are marked in the map.

\subsection{Age and distance distributions}
\label{agedist}

\begin{figure*}
\vbox{\psfig{figure=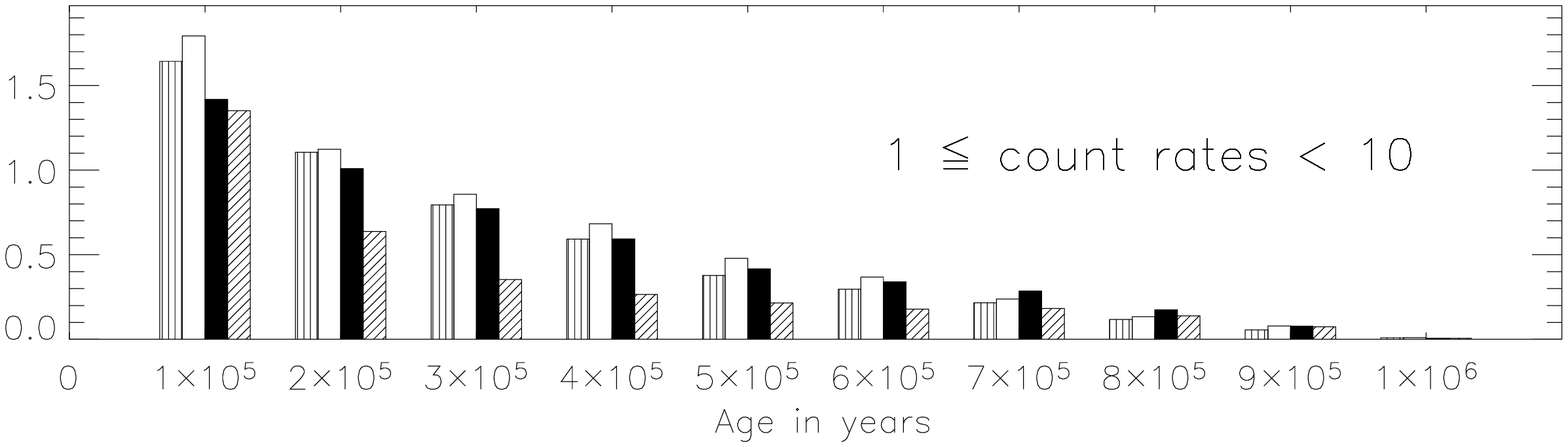,width=17cm}}
\vbox{\psfig{figure=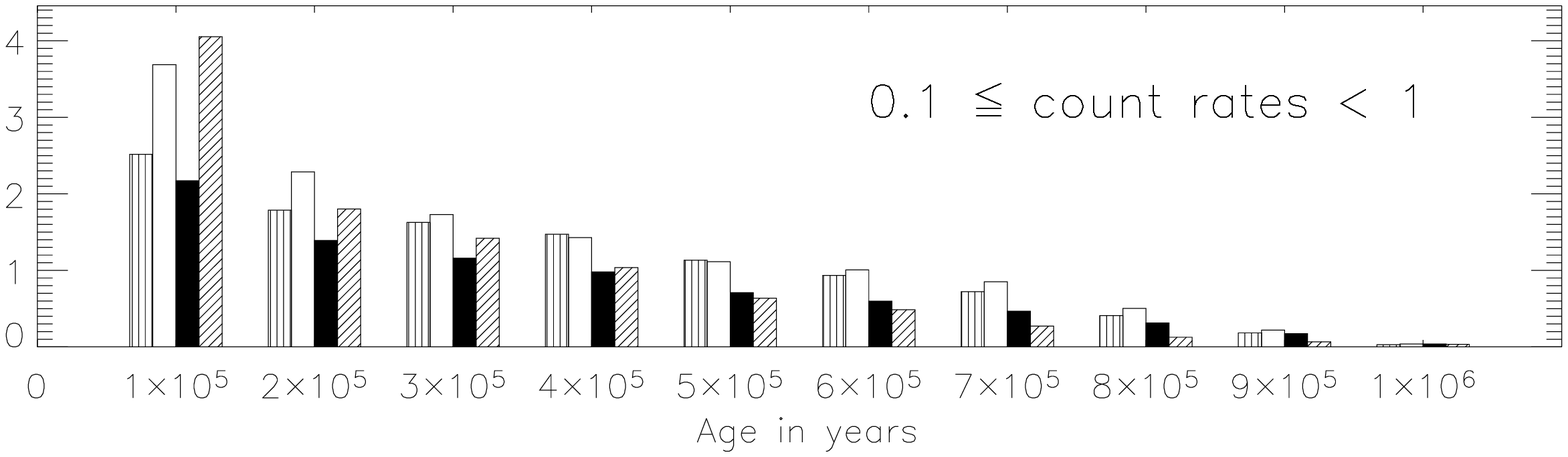,width=17cm}}
\vbox{\psfig{figure=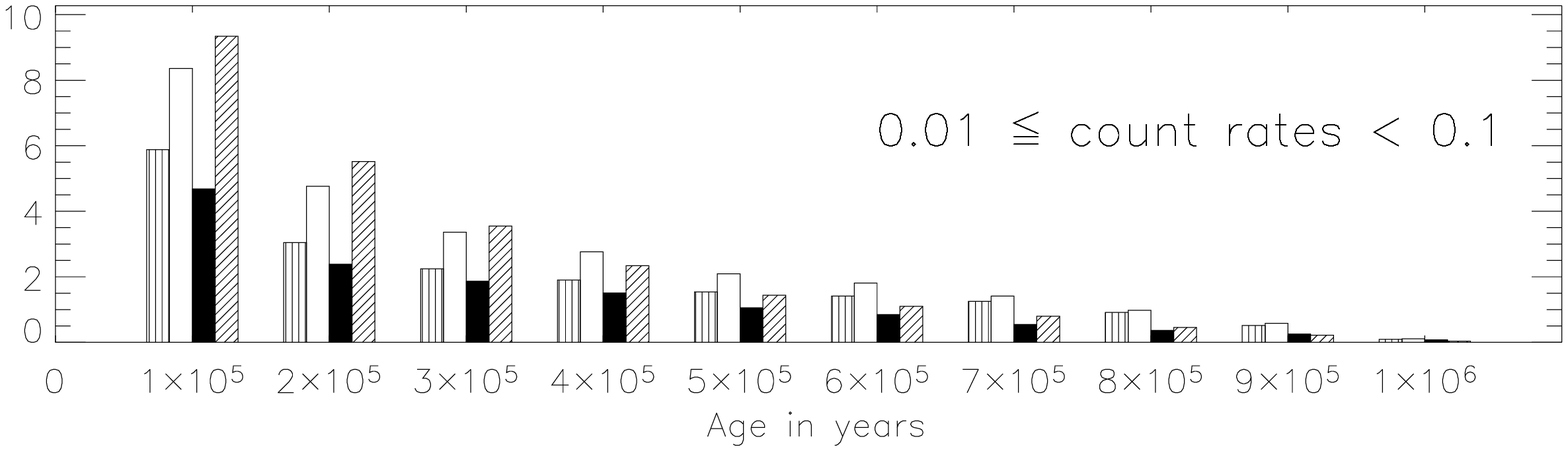,width=17cm}}
\caption[]{In this figure the age
histograms of observable INSs with thermal X-ray emission are plotted.
Three different ROSAT count rate intervals represent very bright,
bright and moderatly bright ROSAT X-ray sources
(note the RASS Bright Source Catalogue limiting count rate of
$0.05$~cts~s$^{-1}$; \citealt{Voges1999}).
Different bars correspond to  different variants of our model.
The bars with vertical stripes indicate the results of the old population
synthesis model with a Gould Belt radius of 500~pc
(Paper II \nocite{p04}, see also Sec.~\ref{ini}),
the white bar correponds to the population synthesis applying
the new inital spatial distribution (Sec.~\ref{ini},
reference $\log$~N~--~$\log$~S curve in Fig.~\ref{fig:ini_xyz}
to Fig.~\ref{fig:massISM}).
Results obtained in the frame of the new analytical ISM model
are shown with black bars, and those obtained with the Hakkila ISM model are
represented by  bars with diagonal stripes
(see Sec.~\ref{ism} and Fig.~\ref{fig:massISM}
for the corresponding $\log$~N~--~$\log$~S curves).
Please note  different y-scales in different panels.
{\bf For an updated version of this figure see Fig.~\ref{fig:Coragehist} in the Appendix~\ref{Erratum}.}
}
\label{fig:agehist}
\end{figure*}

\begin{figure}
%\vbox{\psfig{figure=Disth75h71Pana5kPhak5k_0_1bw.eps,width=8.0cm}}
%\vbox{\psfig{figure=Disth75h71Pana5kPhak5k_1_0bw.eps,width=8.0cm}}
%\vbox{\psfig{figure=Disth75h71Pana5kPhak5k_2_1bw.eps,width=8.0cm}}
\vbox{
\hbox{
\psfig{figure=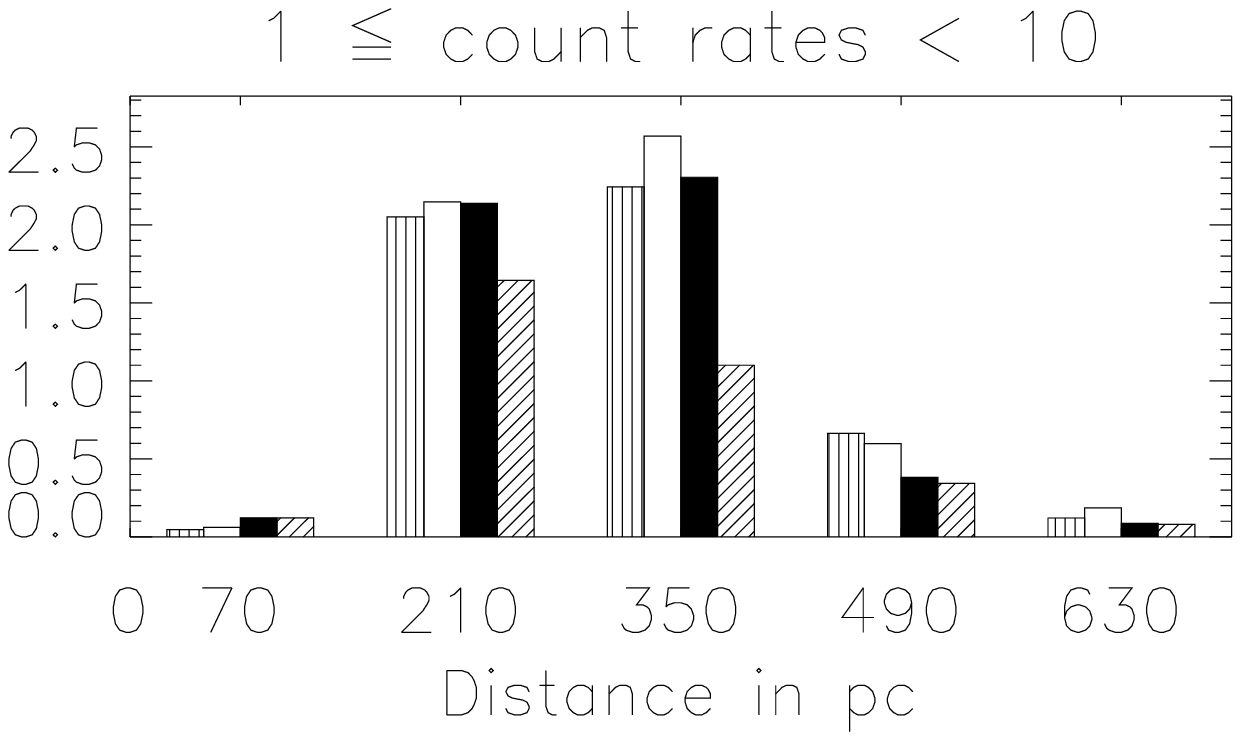,width=8.5cm}
}
\hbox{
\psfig{figure=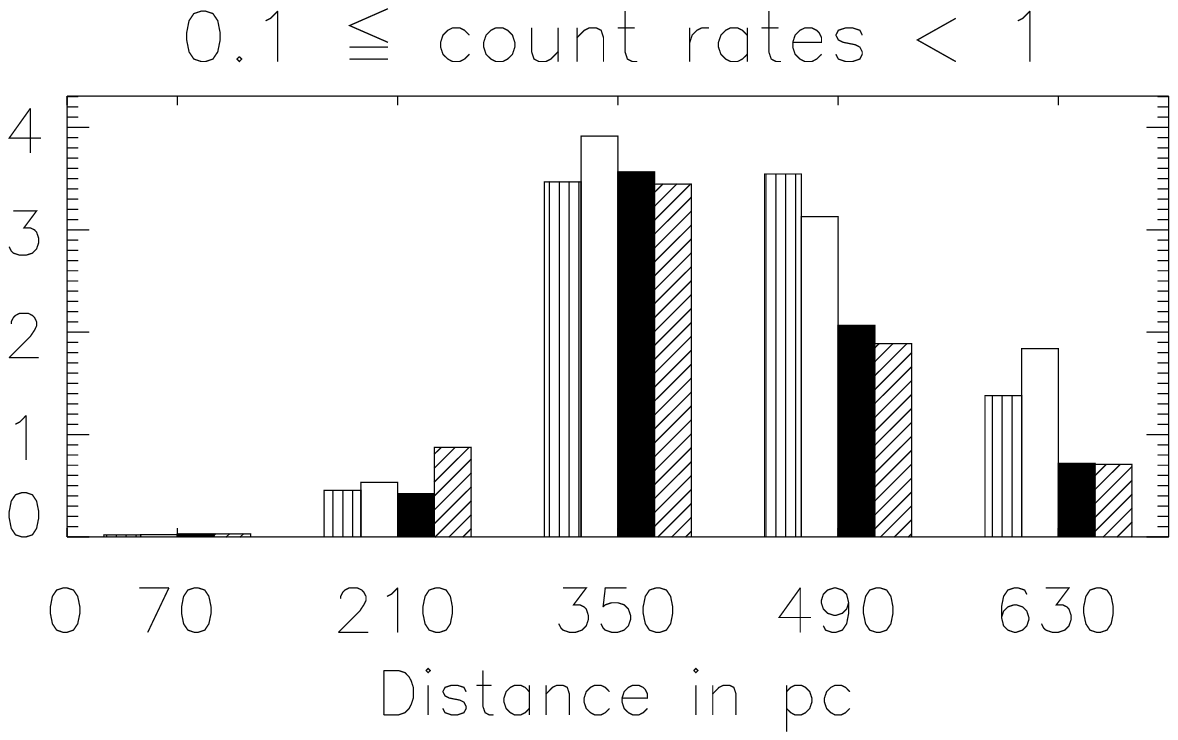,width=8.5cm}}

\hbox{
\psfig{figure=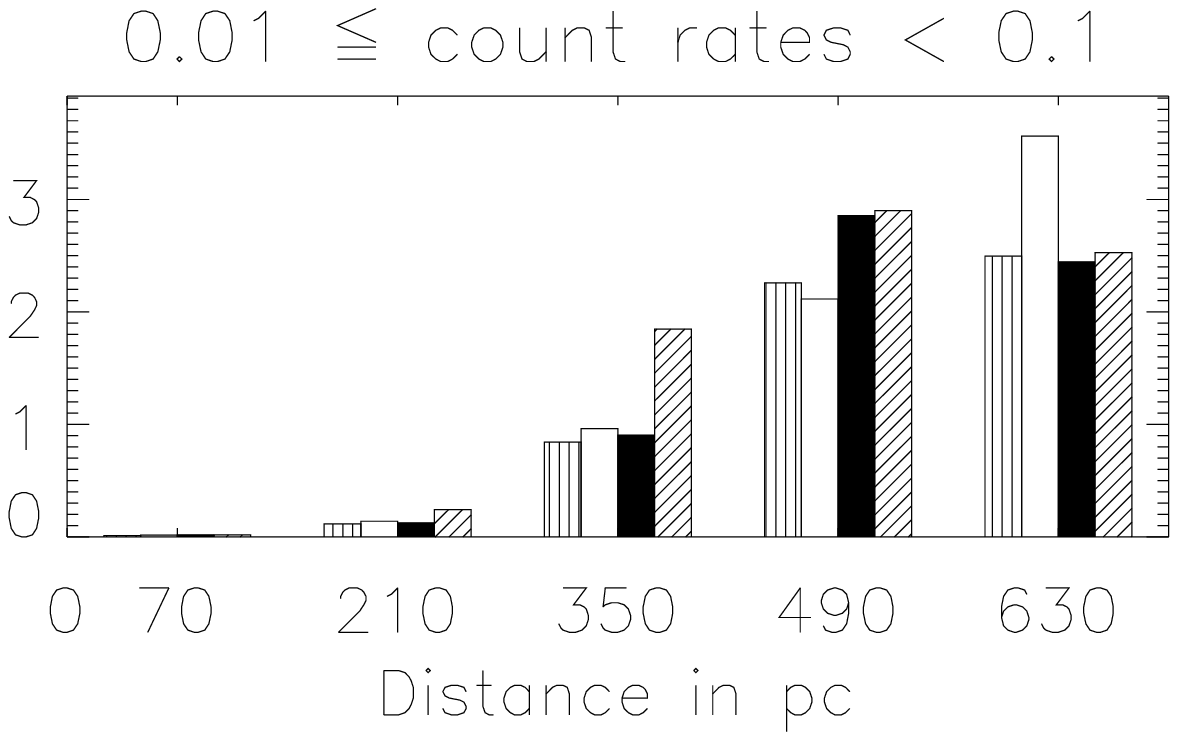,width=8.5cm}
}
}
\caption[Distance histograms]{The distance histograms of the
observable INSs with thermal X-ray emission are shown for three
different ROSAT count rate intervals.
Different bars correspond to different variants of our model,
and have the same pattern as in Fig.~\ref {fig:agehist}.
Please note different y-scales in different panels.
{\bf For an updated version of this figure see Fig.~\ref{fig:Cordisthist} in the Appendix~\ref{Erratum}.}
}
\label{fig:disthist}
\end{figure}

In this paper we introduce age and distance distributions of
close-by NSs as predicted by our PS model. 
These distributions can help to illustrate better
the properties of both discovered and still elusive ICoNSs.
Figures~\ref{fig:agehist} and \ref{fig:disthist} show the computed age and
distance distributions for NSs with ages up to $10^6$ years with
bin size $10^5$~years, and distances up to 700~pc with bin size of
140~pc, calculated for different model assumptions ( the two variants
of the initial spatial distribution of NSs, and the three models of ISM
distributions).

We start by presenting only the general picture and refer to
Sect.~\ref{sec:compaged} for a comparison of the impacts the
different PS scenarios have on our results.
Fig.~\ref{fig:agehist} confirms the natural expectation that the
youngest neutron stars contribute most to the observable object
fraction in all flux ranges as they are both brighter and hotter.
The radial distribution for different count rates is presented in Fig.~\ref{fig:disthist}. 
While for large
fluxes most sources are situated in the region  $\sim 200$~-~$
400$~pc (ie. in the Gould Belt), for  fainter objects the picture
is different. At fluxes below $\sim 0.1$~cts~s$^{-1}$ one expects to
see mostly sources behind the Gould Belt.
These general features are very important, as they indicate that
new, still unidentified ICoNS are expected to be young objects
behind the Gould Belt. We discuss this proposition below in
Sec.~\ref{search}.

\subsubsection{Comparison of the distributions for the different model
modifications} \label{sec:compaged}
The population synthesis models used to obtain the results
presented in Fig.~\ref{fig:agehist} and Fig.~\ref{fig:disthist}
differ on the one hand in the initial progenitor distribution and
on the other hand in the adapted ISM model. As it is already evident
from the $\log$~N~--~$\log$~S curve in Fig.~\ref{fig:ini_xyz} the predicted
number of observable NSs is generally higher in the case
of the new initial progenitor distribution compared (white bars)
to the old one (vertically striped bars). The decrease in number from the youngest to the oldest sources for
both progenitor distributions is similar. For sources with count rate
$<$~1~cts s~$^{-1}$ the new initial spatial distribution results in larger
number of young sources. These are objects behind the Gould Belt
in OB associations. Outside the Gould Belt the new distribution is more
compact than the old one: most of associations are closer than 2
kpc (in the old model NSs were uniformly distributed in
the disc up to 3 kpc).

As the next step we compare the results for the old simple ISM model
(white bars in Fig.~\ref{fig:agehist} and Fig.~\ref{fig:disthist}) with
those obtained using the new analytical model (black bars).
In both cases we use the new inital progenitor distribution.
In general, calculations with the old ISM model produce more observable sources
(which can also be seen in the $\log$~N--$\log$~S curves in Fig.~\ref{fig:massISM}).
{\bf The following detailed comparison of the analytical and Hakkila ISM model for specific directions is obsolete after correcting for the bug in our population synthesis code and can be neglected. The general picture however remains unchanged.}  
This difference is
more clearly pronounced for faint sources with count rates below 1 cts s~$^{-1}$ and
ages below $3 \times 10^5$~years.
The X-ray emission of these objects must be more absorbed in the new analytical ISM model to explain this effect.
Therefore, these young faint sources have to be located on directions
which correspond to larger hydrogen column densities  in the new
analytical ISM model. Indeed, there are several small regions of higher
column density in the Galactic plane, and some of them
hinder the view to star associations, e.g. at $l=215^{\circ}$.

The distance histograms at different count rates at low distances
(e.g. up to the third distance bin, $[ 280, 420 ]$~pc) show
no difference or only slight reduction comparing
the resulting numbers from applying the old ISM with those obtained by using
the new analytical ISM model.
At larger distances
in the distance histograms 
there are fewer observable NSs
in the case of
the new analytical model in comparison with the old one,
e.g. in the last distance bin, $[ 560, 700 [$~pc,
in case of the faintest sources.
These differences are evident from the properties of the new
analytical ISM model.
First, due to the
inhomogeneities in the new analytical ISM model there are regions of higher
column density compared to the old model. 
Second, the ``average'' column density (without apparent ISM clumps) in the
Galactic plane is less for the new model at distances below 450~pc. 

The lower average N(H) value at
relatively close distances can explain the comparable or even slightly
increased predicted number of observable neutron stars even with present ISM clumps.
The ISM clumps are however responsable for the number reduction in the
right part of the distance histograms. For example there is strong
absorption towards the direction $l=100^{\circ}$, $b=-5^{\circ}$ relatively
close to the Cygnus-Cepheus OB-associations region. The predicted number of observable
NSs is reduced there. Since the OB associations towards this
direction of sight are at distances larger than 500~pc (e.g. Cyg OB7 at
740~pc, Cep OB2 at 700~pc and Cep OB3 at 900~pc, Cam OB1 at 900~pc) the
effect is pronounced mostly at faint count rates (objects relatively far
away).

Now we compare the predicted numbers calculated using  the analytical ISM
model (black bars) and the Hakkila extinction model (diagonal
stripes).
Here, we note again that up to a 
distance of 230~pc both ISM models
are almost the same.
From 240~pc the extinction study by \cite{Hakkila97}
is applied, and so starting from this point
the absorption in this more inhomogeneous model is usually much larger than
in the case of the analytical model which is characterized by slower and more homogeneous 
increase of the column density.
Only at distances
around 450~pc do the hydrogen column densities of the analytical ISM model
become larger than those of the Hakkila ISM model in most regions.
For example, the Hakkila ISM model has in general lower hydrogen column
densities at $|b|>10^{\circ}$ for distances larger than 400~pc.
Extinction is also low in this ISM model from $l=230^{\circ}$ to $l=285^{\circ}$ in the
Galactic plane (all distances), where differences of up to an order of
magnitude
can be reached in comparison to the analytical ISM model.\\
In agreement with the $\log$~N--$\log$~S curves of Fig.~\ref{fig:massISM}
the Hakkila NS number of bright sources is slightly less than
in the case of the analytical model,
both in the age and the distance histograms.
This slightly lower observable NS number is due to
larger absorption at closer distances.
For count rates below 1~cts s~$^{-1}$ the application of the Hakkila ISM leads to a higher number of observable NSs, especially of young age.
The lower absorption at large distances explains these higher observable number of ICoNSs.
For the discussed distance range, however, the distance histograms indicate only small differences for sources fainter than 1~cts s~$^{-1}$.

\subsection{Resultant search strategies}
\label{search}

The main aim of this study is to make advances in the
strategy for searching for new ICoNSs. Based on the results
presented above we can suggest some modifications in the approach
to look for new candidates. According to what we discussed in the
previous sections, new candidates expected to be identified at
ROSAT count rates $<0.1$~cts~s$^{-1}$ should be young objects born
in rich OB associations behind the Gould Belt. Most of the recent
studies \cite{Agueros2006,Chieregato2005,Rutledge2003}
looked for new candidates far from the Galactic plane. It seems
that this is not very promising. Our results indicate that ICoNS
should be searched in directions of OB associations such as Cyg
OB7 and Cep OB3. On average, new candidates should be slightly
hotter than the known ICoNSs, as they are younger.

The absence of sources in the second Galactic quadrant
\citep{Motch2007}
can be naturally explained taking into account the properties of
close-by OB associations. Most of the associations (Per OB2, Cas-Tau,
$\alpha$ Per, Cep OB6) marked in this empty region in Fig.~3
from \cite{Motch2007} have ages which do not favor the appearance
of young  observable  NSs at present times. They are too young
($<$~7~Myrs) or too old ($>$~25~Myrs), see Table~1 above. As noted
above, the strong contribution of the Cygnus-Cepheus region to the
expected neutron star number, shown in map Fig.~\ref{fig:map}, comes
especially at low count rates. Thus, we expect that sources with
50$^\circ$~$<$~l~$<$~200$^\circ$ would be identified having
lower fluxes  than the known sources, since
they come from
OB associations further away.

Considering sky coverage the ROSAT All Sky Survey
(RASS) is currently the best choice to look for new ICoNSs in the 
Cygnus-Cepheus region which is, according to our presented results, the most promising region. However, the relatively
large positional error circle of ROSAT usually includes many
possible optical counterparts, especially at these low Galactic
latitudes. Furthermore one has to exclude variable X-ray sources
to find ICoNSs. In this respect the recently published
XMM-$Newton$ Slew Survey (EPIC with $medium$ filter, e.g. \citealt{Esquej2006}) may
become an important database. Currently this survey  (0.2--12~keV
band) covers roughly $~ 15$~\% of the sky, and below 2~keV its
sensitivity is comparable to that of the RASS
\citep{Freyberg2006}. The sensitivity is however strongly
inhomogeneous perpendicular to the slew direction. The astrometric
accuracy is around $17\arcsec$ (90\% confidence level), but in few
cases an error of $1\arcmin$ due to an attitude problem has been
reported\footnote[1]{http://xmm.esac.esa.int/external/xmm$\_$science/slew$\_$survey/xmmsl1$\_$ug.shtml}.
In conjunction with the RASS, the XMM-$Newton$ Slew Survey with its increasing coverage can be
used to identify non-variable X-ray sources and eventually
improve their positional accuracy. With the planned expanded
energy range down to 150~eV (M.Freyberg, pers. comm.) the soft
X-rays emitted by ICoNSs could be detected. 
However, the
XMM-$Newton$ Slew Survey has  strongly inhomogeneous
sensitivity, and the RASS is not deep enough for the expected new
faint sources. Positions of both can be too inaccurate in the
highly populated Galactic plane. Pointed observations by
XMM-$Newton$ with its high sensitivity and by $Chandra$ with its
superb positional accuracy may solve these problems for particular directions.

\subsection{Outlook}
\label{outlook} 
Whether the calculated $\log$~N--$\log$~S curves are
realistic descriptions of the ICoNS numbers also at low count rates
could be best tested with the help of an All Sky Survey like that with the planned eROSITA\footnote[2]{www.mpe.mpg.de/erosita/MDD-6.pdf}. This
experiment  is expected to have an angular resolution of
$<25\arcsec$ and to be 30~times more sensitive than the RASS
\citep{PredehlEROSITA}. eROSITA's CCD technology is based on an
improved concept of the successful XMM-$Newton$ EPIC-pn CCDs.
While the noise at low X-ray energies is expected to be much lower
compared to the EPIC-pn CCDs \citep{Meidinger2006} the influence
of the planned eROSITA filter actually reduces the sensitivity to or below
that of ROSAT in case of energies lower than 0.2~keV (P.~Predehl,
pers. comm.). Using a preliminary eROSITA response
matrix one can estimate that detections of soft blackbody spectra
are possible with approximately one eighth of the RASS count rate
after the planned 4 years of survey observations. Thus, in
principle, sources could be expected to reach count rates down to
$\log$~S$\approx -2.7$ (using 0.017~counts~s$^{-1}$ as the RASS
limit). This can result in a significant increase in the number of
currently known ICoNSs, up to an order of magnitude. However, the
actual number of identified NSs will depend strongly on the
finally reached positional accuracy, especially in the Galactic
plane. For finding new sources of the Magnificent Seven type it will be advantageous to have X-ray positions with $\sim 1\arcsec$ accuracy to avoid confusion with faint extragalactic objects \citep{trumper2006}.

Besides the standard approach based on X-ray/optical wavelengths we mention below  other possibilities to look for ICoNSs. They are not directly related to our PS results. Keeping the restrictions of our model in mind, however, they may be important in terms of finding new ICoNSs.

Recently,
\cite{Crawford2006} found 56 well-determined EGRET error boxes devoided of radio pulsars.
Estimates of the
number of young NSs which can manifest themselves as
$\gamma$-ray sources show that many of the 56 sources can be related to
these compact objects,
but a significant fraction of them should not appear as standard
radio pulsars \citep{Harding2007,Gonthier2007}.
Depending on the $\gamma$-ray emission model the fraction of Geminga-like objects
can reach  between 30~$\%$ and 90~$\%$ of $\gamma$-ray pulsars according to \citet{Harding2007}.
They can be detectable in soft X-rays as {\it coolers}, elusive in the radio
band\footnote[3]{Geminga itself was claimed to be detected in radio \cite{Malofeev1997},
 but the ``second
Geminga'' -- 3EG J1835+5918 -- is observed only in $\gamma$ and in
soft X-rays. \protect\\ \hspace{5cm} \protect\\ \hspace{5cm} \protect\\}. New GLAST and AGILE observations can provide much
smaller error boxes, so soon it could be possible to identify
sources if they are ICoNSs. It would be fascinating to find more
``$\gamma$-ray selected'' {\it coolers}. Of course, it is
important to make a joint population synthesis for thermal and
non-thermal X-ray sources, but this task is out of the scope of
this paper.

Another possibility to find new ICoNSs is
to search for (un)bound compact companions of OB runaway stars.
More than one hundred OB runaway stars are known in a 1 kpc region around the
Sun \citep{Zeeuw1999}.
In comparison with typical OB stars they are characterized by large spatial velocities or/and by large distances from the Galactic plane. Two main
origins of these large velocities are currently discussed: dynamical interaction and expulsion of a companion in a close binary system (e.g. \citealt{Blaauw1961}). The latter case is
interesting for the discussion of search for new close-by cooling NSs.

A binary can survive after the first SN explosion in, roughly,
10-20 \% of cases (e.g. \citealt{Popov2006}). Then a runaway system
consisting of an OB star and a compact object (most probably a
NS) is expected. A young NS can appear as a radio pulsar. \cite{sayer1996} and
\cite{philp1996} searched for radio pulsar companions of $\sim 40$
runaway OB stars. Nothing was found. This result is consistent
with the assumption that in less than 20\% cases OB stars have
radio pulsar companions. Still, it is interesting to speculate
that runaway massive stars can have ICoNS as companions. Then a
companion can be identified as a source of additional X-ray
emission if a NS is younger than
$\sim10^6$~yrs. It is expected to have a thermal component
with $T\sim 50$~-~200~eV and $L\sim
10^{30}$~--~$10^{32}$~erg~s$^{-1}$ in most standard scenarios of
cooling (see \cite{bgv2004,page2006,Page2007} and
references therein). Our simple estimates show that a cooling NS in a
binary with an OB star can appear in the propeller stage. In this
case additional energy output non-related to the surface thermal
emission can be expected.

As mentioned above, in our calculations very few NSs can reach
high Galactic latitudes. However, in our model we do not take into
account that progenitors themselves can be high velocity objects
(runaway stars or so-called hyper-velocity stars,
\citealt{Brown2005}). In this case a NS can be born far away from the
Galactic plane. This possibility is important in connection with
the recently discovered source, Calvera \citep{Rutledge2007}, 
which could have had a high velocity progenitor.
Recent strict bounds on the radio emission from Calvera \citep{hsr2007} make it probable that this source is similar to the {\it Magnificent Seven}.

Hyper-velocity stars \citep{Brown2007} are expected to originate mainly
from the Galactic center . They acquire large velocities after
binary disruption in the field of the supermassive Galactic black hole. 
Another possibility is the ejection from young star clusters \citep{Gvaramadze2007}.
Initial
velocities are about 1000~--~3000~km~s$^{-1}$. In 25 Myr (maximum
lifetime of a NS progenitor) a hyper-velocity star can travel up
to 50 kpc, more than enough to explain Calvera. However, the
number of such stars is not high. The total rate of production of
hyper-velocity stars of all masses is between $10^{-3}$ and
$10^{-6}$~yr$^{-1}$ \citep{Brown2007}. If we
take this rate as $\sim 10^{-6}$~yr$^{-1}$ for NS progenitors then one can expect a
few ICoNS on very high Galactic latitudes. In this case it is not
very probable to find one Calvera in $\sim 10$~kpc from us. Still,
this possibility cannot be excluded.
The high velocity tail of runaway stars corresponds to 200-300
km~s$^{-1}$. In 25 Myr a NS progenitor can reach $z\sim 5$~kpc
above the Galactic plane. Potentially, it is enough to explain
Calvera, but detailed modelling is desirable. 

Of course population synthesis models in general,
and the one we use in particular, cannot predict new types of
sources, since our results do not include populations of coolers absolutely
different from those assumed in the model. For example, Calvera can be an evolved
version of the Cas A central compact object (CCO, for more information on CCOs see e.g. \citealt{Kaspi2006}). 
Our model does currently not
include objects like this (small emitting area, etc.). In general, the
differences in properties between CCOs and close-by coolers indicate that we currently
do not understand initial properties (and perhaps, evolution) of
NSs well enough.  

A joint population synthesis of all known types of NSs would be useful. The
presented population synthesis model cannot take into account all
interesting possibilities related to evolution of cooling NSs.
So, as usual, surprises are not excluded.

\section{Conclusions}
\label{concl}

In this paper we presented our new more advanced model for population
synthesis of close-by cooling NSs. 
The two, sligthly different, mass distributions we consider result in nearly the same observable number of NSs.
Detailed treatment of the initial spatial
distribution of NS progenitors and a detailed ISM structure up to 3 kpc allow us to discuss the strategy to look for new ICoNS.
Our main results in this respect are the following: new candidates are
expected to be identified behind the Gould Belt, in directions to rich OB
associations, in particular in the  Cygnus-Cepheus region; new candidates,
on average, are expected to be hotter than the known population of ICoNS.
Besides the usual approach (looking for soft X-ray sources), the search in 
'empty' $\gamma$-ray error boxes or among run-away OB stars may yield new X-ray
 thermally emitting  NS candidates.

\begin{acknowledgements}
We thank D. Blaschke, H. Grigorian, and D. Voskresensky for data on cooling
curves and discussions; A. Mel'nik for discussion of properties of OB
associations; R. Lallement for the sodium data; and A. Pires for discussions about the ISM model.  The authors thank the anonymous referee for careful reading of the manuscript and the suggested improvements.

S.B.P. was supported by INTAS and Dynasty foundations.\\
This research has made use of SAOImage DS9, developed 
by Smithsonian Astrophysical Observatory; the SIMBAD and VizieR databases,
operated at CDS, Strasbourg, France; and NASA's Astrophysics Data System Bibliographic Services.
\end{acknowledgements}

\appendix
\section{Erratum}
\label{Erratum}
We report in the following on some details regarding the corrected  results obtained by the upgraded version of our population synthesis code.
All our main conclusions presented above remain valid.

\subsection{Comparison of $\log$~N~--~$\log$~S distributions for different model modifications}
\label{lognlogsC}
\begin{figure}[ht]
\hspace{-1cm}
\psfig{figure=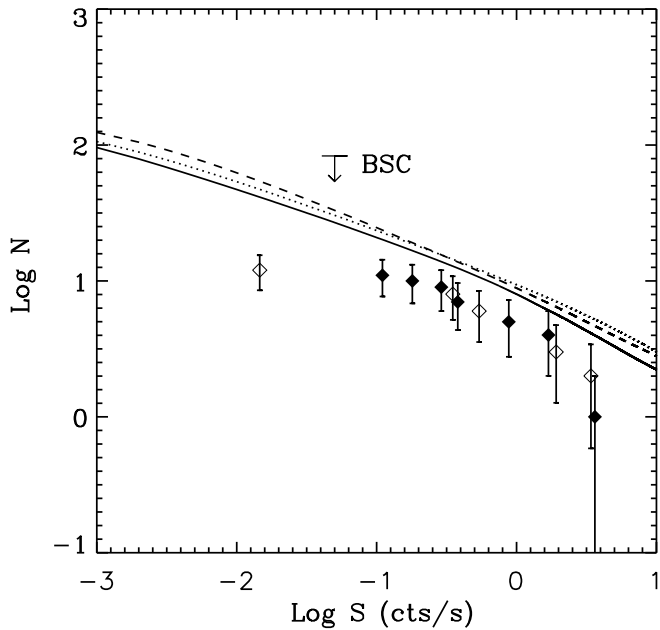,width=10cm}
\caption[]{Log N~--~Log S for different X-ray absorbing ISM models.
This figure is the corrected version of Fig.~\ref{fig:massISM}
All curves are plotted for the new initial spatial distribution, the old mass spectrum and new ISM  element abundances.
Solid curve: old, simple analytical ISM model as, e.g., in
Paper III \nocite{pgtb04}.
Dotted curve: new improved analytical ISM model (corrected);
dashed curve: Hakkila ISM model (corrected).}
\label{fig:massISMupd}
\end{figure}

The $\log$~N~--~$\log$~S curves in Fig. 3 and Fig. 4 of the original paper remain unchanged.
The $\log$~N~--~$\log$~S curves for the new analytical as well as for
the Hakkila ISM model in Fig. 5 of the original paper are updated in Fig.~\ref{fig:massISMupd}.
The corrected curves for both ISM models are now situated $\approx 0.3$\,dex above the observational points. 
The differences between the Hakkila ISM and the improved ISM analytical model are smaller than those obtained by the original PS-code. However, as before, the application of the Hakkila ISM model results in lower $N$ for high count rates than obtained by using the analytical ISM model, and in higher $N$ for low count rates.
Comparison of our new results with observations of bright, cooling NSs indicates that the model overpredicts the number of NSs by roughly a factor of two for both ISM models. The possible reasons for this discrepancy are an inadequate treatment of the NS birth rate or of their thermal evolution, or other, not yet in this paper investigated properties like atmospheres, magnetic fields, or statistical fluctuations. 
Birth rates of neutron stars are highly uncertain, especially at larger distances (see, e.g.,  the recent discussions by \citet{Keane2008} and \citet{Ofek2009}).   
However, since probably nearly all the observed XTINSs originate from the Gould Belt 
we have to discuss the \emph{local} birth rate in the frame of the differences between our PS model predictions and the observational measurements. 
As discussed in the paper (Sect. 2.1), we 
adopted a birth rate of 27~Myr$^{-1}$ up to a distance of 500\,pc \citep{g2000,tammann} and of 270~Myr$^{-1}$ for the whole distance range from 0 to 3000 pc \citep{tammann}.
The supernova rate ranges from 17~Myr$^{-1}$ to 27~Myr$^{-1}$  in the entire Gould Belt \citep{Grenier2004}. Only 75\% to 87\%  of the core-collapse supernova produce neutron stars \citep{Heger2003}.
Thus, the birth rate in the Gould Belt we chose can be overestimated.
However, a factor of two in the {\emph{local}} birth rate uncertainy seems to be unlikely.

As mentioned in the original paper, the main conclusions about different sets of cooling curves presented in Paper III do not change.
The chosen set of cooling curves in this paper actually represents the best
choice of the set from Paper III. All other cooling curves from Paper III would
result in an even higher $\log$~N. 
It is beyond the scope of this erratum
to identify a new cooling curve set
that could lead to a $\log$~N~--~$\log$~S curve in better agreement with observational points.

We further note the possible effect of statistical fluctuations on the uncertainty of the observed $\log$~N~--~$\log$~S curve, since the overall number of (young enough) neutron stars in the Gould Belt is small. We will evaluate this effect in more detail in a future paper.

\subsection{Sky maps}
\label{mapsC}
\begin{figure*}[t] 
 \centering 
 \includegraphics[width=0.9\hsize]{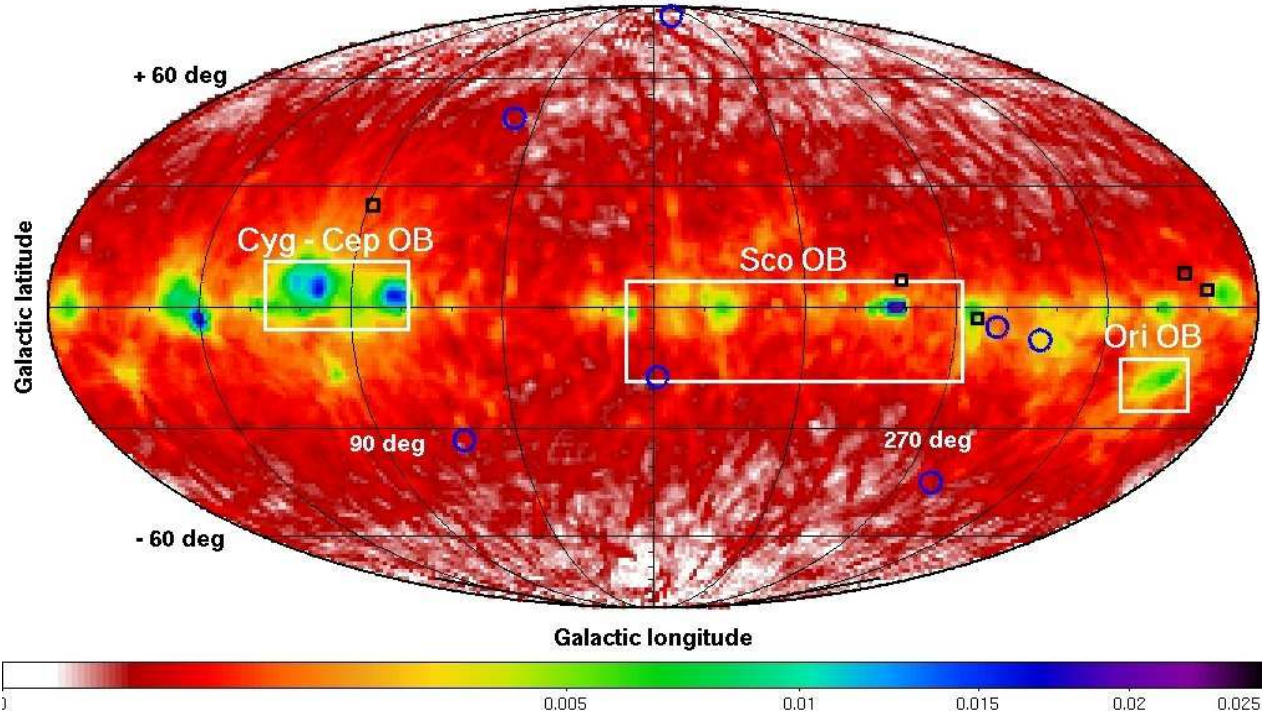}
\caption{This map is the corrected version of Fig.6, showing the expected number density of isolated neutron stars with thermal X-ray emission in units of numbers per square degree.
The Galactic map is in Mollweide projection.
Only sources with ROSAT PSPC count rates larger than 0.05~cts s$^{-1}$ are
considered.
Marked in blue are the positions of the Magnificent Seven and
in black the positions of close young radio pulsars with detected thermal X-ray emission. Note that the runs for Figs.6 and 7 on the one hand, and Figs.~\ref{fig:mapErr} and \ref{fig:map2Err} on the other hand, are separate realisations of our code (e.g., radomised birth places); therefore, individual NS "tracks" are not the same.}
\label{fig:mapErr}
\end{figure*}

\begin{figure*} 
 \centering 
 \includegraphics[width=0.9\hsize]{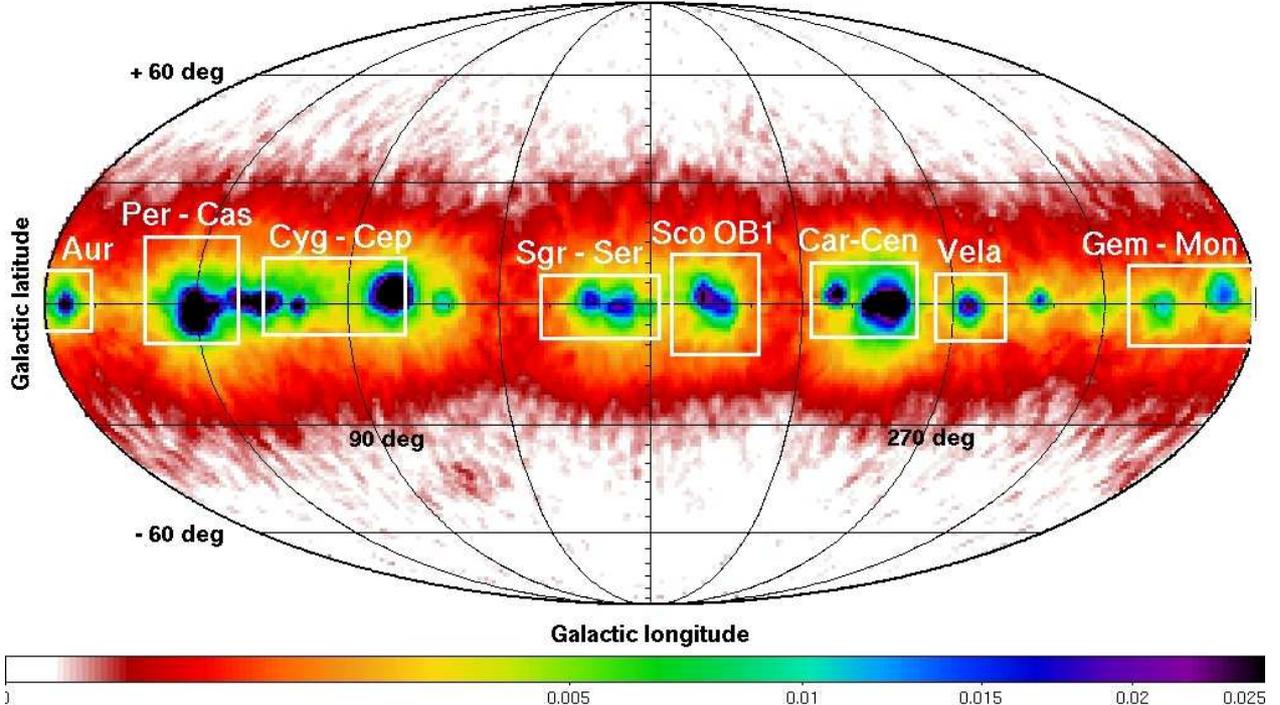}
\caption{This map is the corrected version of Fig.7. Here only faint sources with ROSAT PSPC count rates between 0.001~cts s$^{-1}$ and  0.01~cts s$^{-1}$ are considered.}
\label{fig:map2Err}
\end{figure*}

For completeness we show in Figs.~\ref{fig:mapErr} and \ref{fig:map2Err} the corrected versions of Figs.\,6 and 7 from the original paper. The general picture remains the same, but on average more NSs are expected in accordance to the $\log$~N~--~$\log$~S curve. For bright sources, a subgroup of Sco OB2, Lower Centaurus Crux \citep{Zeeuw1999}, is predicted to have more observable NSs compared to other Sco OB groups, which was not visible in the map obtained with the original version of the PS code.

\subsection{Age and distance distributions}
\label{agedistcor}
Age and distance diagrams obtained with the corrected PS-code show the same main features as those obtained with the old code except that the NS numbers of the analytical and Hakkila ISM model (the last black and diagonal-striped bars in Figs.\,8 and 9 of the original paper) are larger than before, as is expected from the $\log$~N~--~$\log$~S curves.\\

\begin{figure*}
\vbox{\psfig{figure=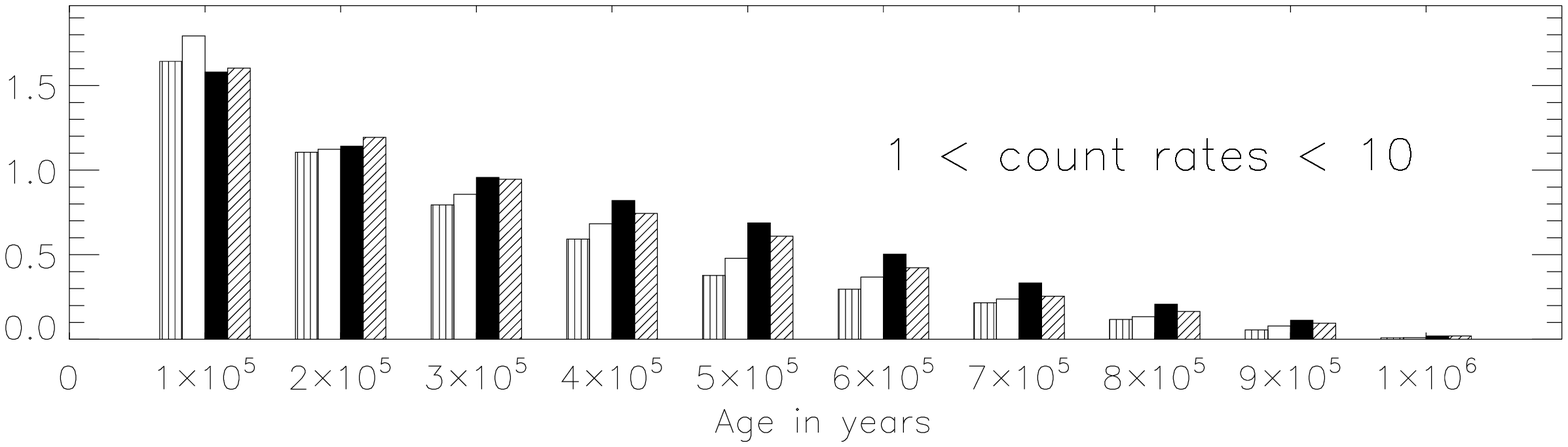,width=17cm}}
\vbox{\psfig{figure=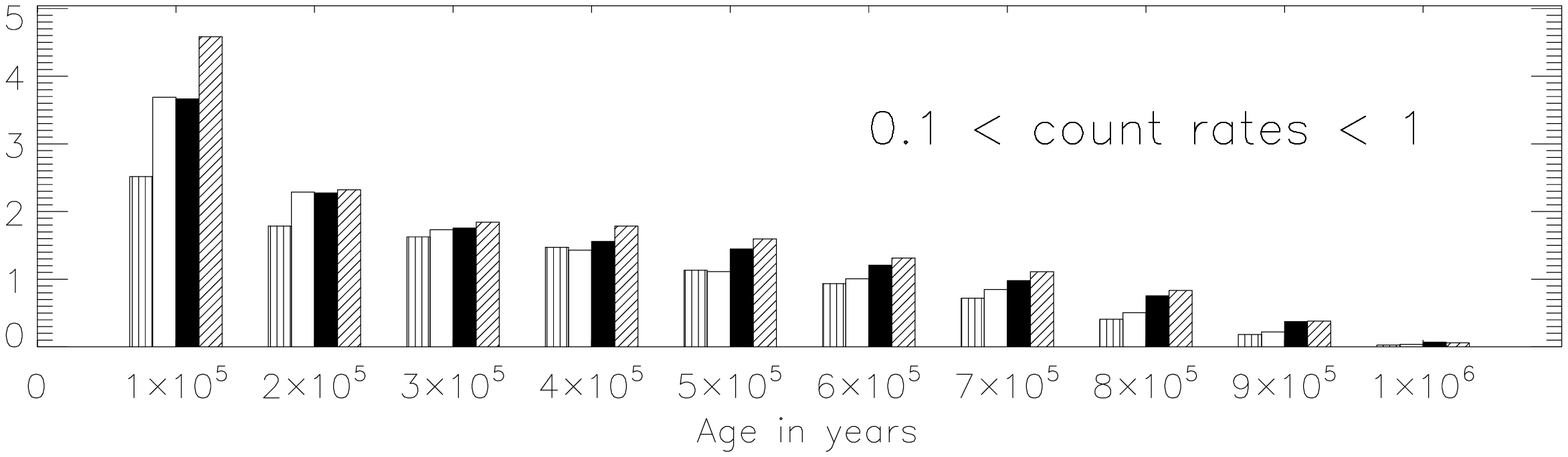,width=17cm}}
\vbox{\psfig{figure=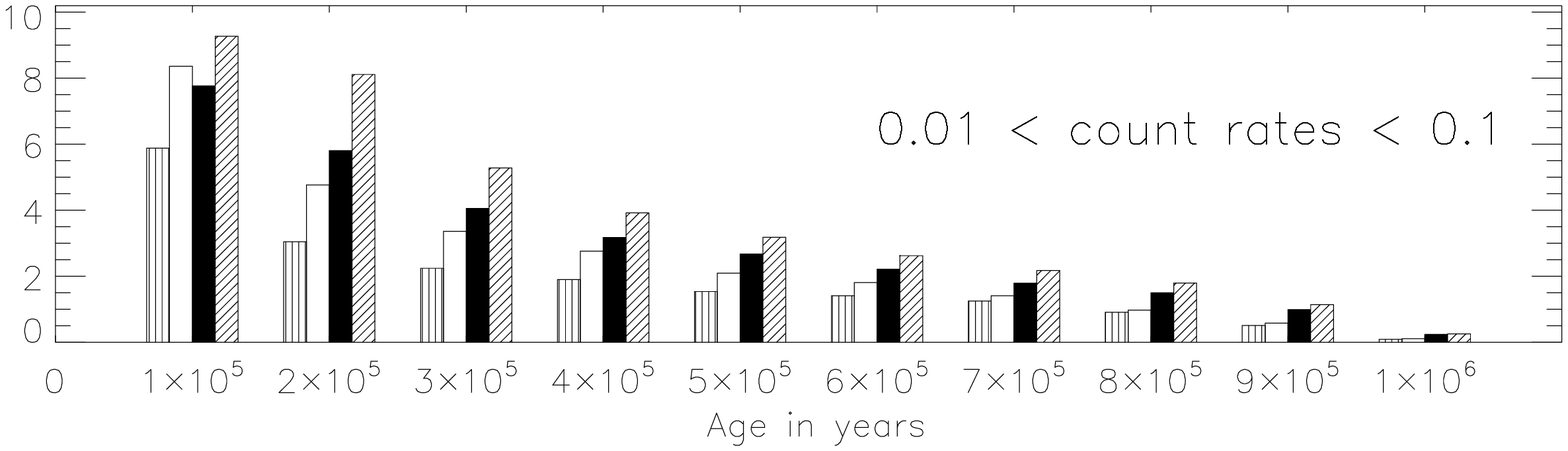,width=17cm}}
\caption[]{This figure is the corrected version of Fig.~\ref{fig:agehist}.
It shows the age
histograms of observable INSs with thermal X-ray emission.
Three different ROSAT count rate intervals represent very bright,
bright and moderatly bright ROSAT X-ray sources
(note the RASS Bright Source Catalogue limiting count rate of
$0.05$~cts~s$^{-1}$; \citealt{Voges1999}).
Different bars correspond to  different variants of our model. 
The same pattern as in Fig.~\ref {fig:agehist} are used:
The bars with vertical stripes indicate the results of the old population
synthesis model with a Gould Belt radius of 500~pc
(Paper II \nocite{p04}, see also Sec.~\ref{ini}),
the white bar correponds to the population synthesis applying
the new inital spatial distribution (Sec.~\ref{ini},
reference $\log$~N~--~$\log$~S curve in Fig.~\ref{fig:ini_xyz}
to Fig.~\ref{fig:massISM}).
Results obtained in the frame of the new analytical ISM model (corrected)
are shown with black bars, and those obtained with the Hakkila ISM model (corrected) are
represented by  bars with diagonal stripes
(see Sec.~\ref{ism} and Fig.~\ref{fig:massISMupd}
for the corresponding $\log$~N~--~$\log$~S curves).
Please note  different y-scales in different panels.
}
\label{fig:Coragehist}
\end{figure*}

%\newpage
\begin{figure}
\hspace{-1cm}
\vbox{
\hbox{\psfig{figure=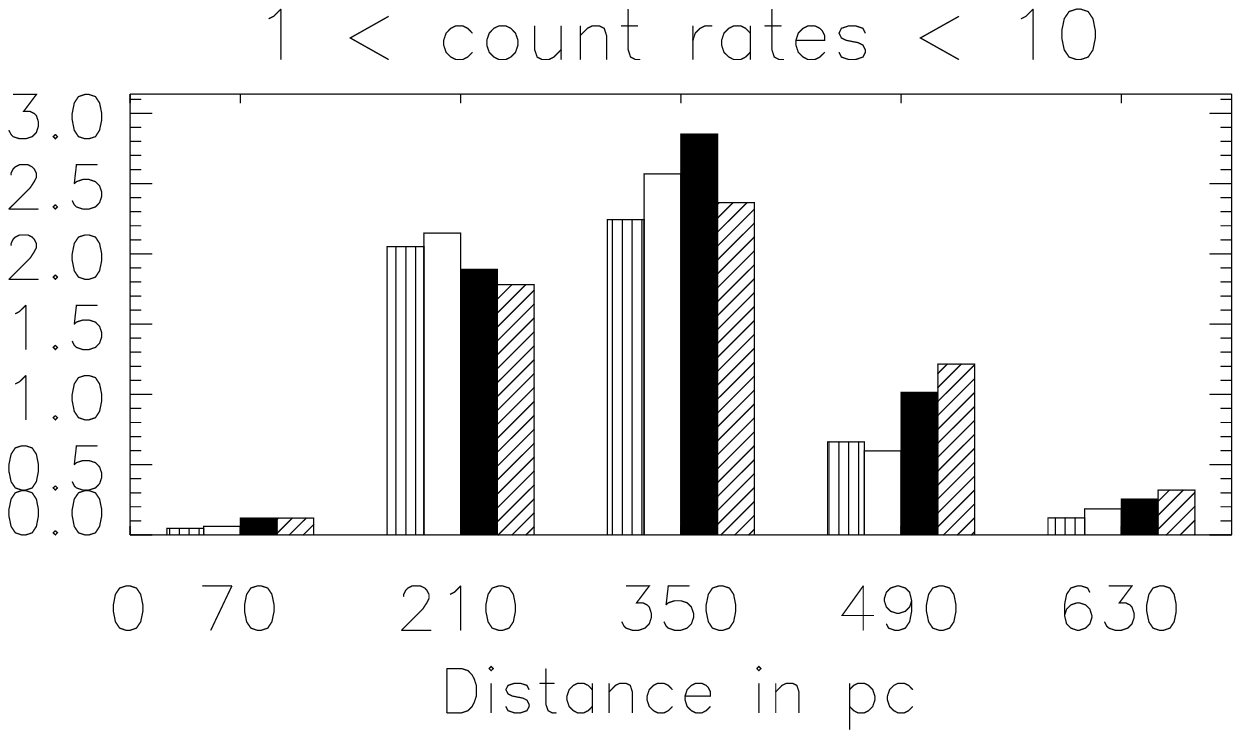,width=9.5cm}}
\hbox{\psfig{figure=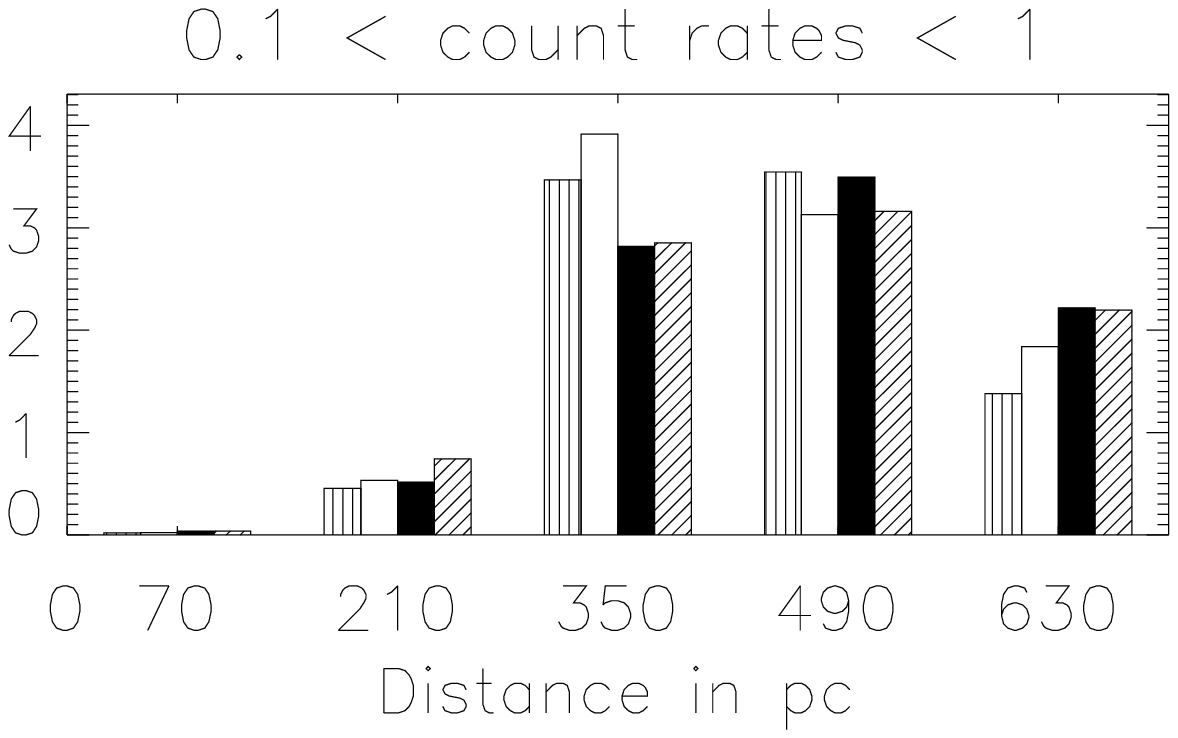,width=9.5cm}}
\hbox{\psfig{figure=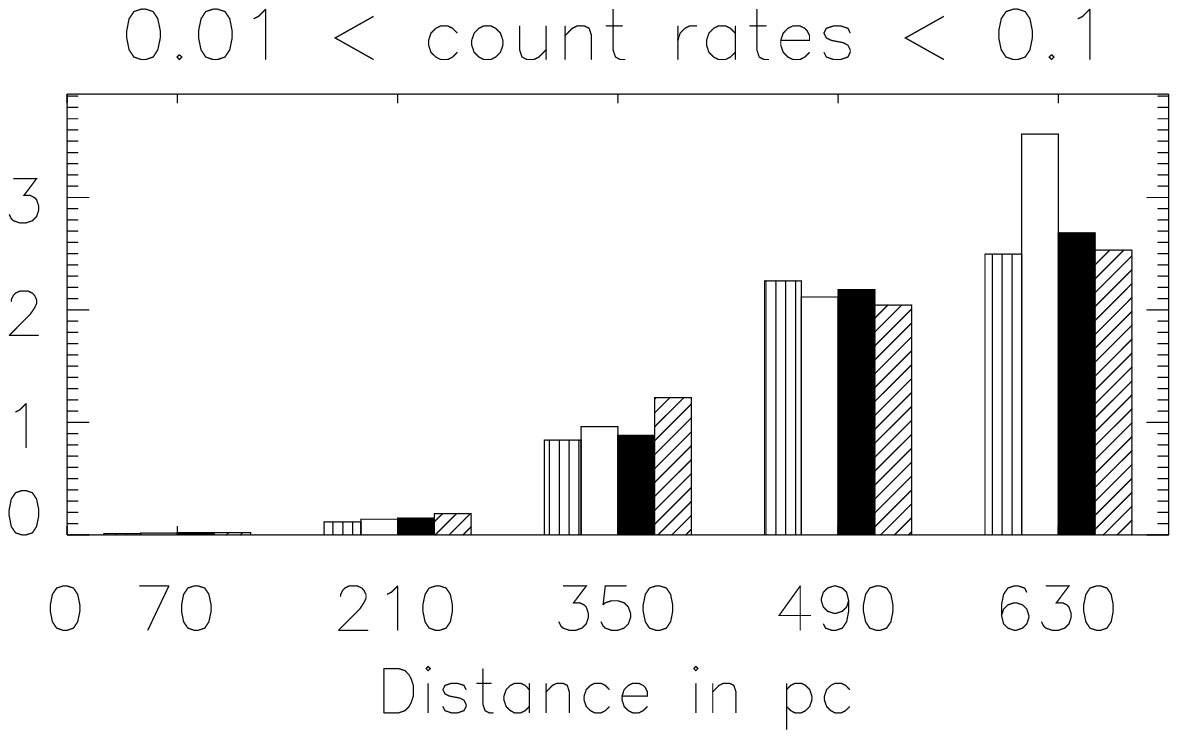,width=9.5cm}}
}
\caption[]{This is the corrected version of Fig.~\ref{fig:disthist}, the distance histograms of the observable INSs with thermal X-ray emission for three
different ROSAT count rate intervals. Different bars correspond to different variants of our model, and have the same pattern as in Fig.~\ref {fig:agehist}. Please note different y-scales in different panels.}
\label{fig:Cordisthist}
\end{figure}
 
\newpage
\clearpage

\end{document}